\documentclass[sigconf, nonacm]{acmart}

\newcommand{\stitle}[1]{\vspace{1ex}\noindent\textbf{#1}}
\usepackage{xspace}
\newcommand{\tree}{B$^S$-tree\xspace}
\newcommand{\ctree}{CB$^S$-tree\xspace}
\newcommand{\bt}{B$^+$-tree\xspace}
\newcommand{\maxkey}{MAXKEY\xspace}

\newcommand{\hide}[1]{}

\usepackage{fancyvrb}
\usepackage[ruled, linesnumbered, vlined]{algorithm2e}
\usepackage[noend]{algorithmic}
\usepackage{subcaption}     
\renewcommand{\algorithmcfname}{ALGORITHM}
\SetKwInOut{Input}{Input}
\SetKwInOut{Output}{Output}
\SetKwInOut{Variables}{Variables}
\SetKwComment{comm}{\hfill$\triangleright$\ }{}

\begin{document}

\title{\tree: A gapped data-parallel B-tree}

\author{Dimitrios Tsitsigkos}
\affiliation{%
  \institution{Archimedes, Athena RC, Athens, Greece}
 }
\email{dtsitsigkos@athenarc.gr}

\author{Achilleas Michalopoulos}
\affiliation{%
  \institution{University of Ioannina \& Archimedes, Athena RC}
 }
 \email{amichalopoulos@cse.uoi.gr}
 
\author{Nikos Mamoulis}
\affiliation{%
  \institution{University of Ioannina \& Archimedes, Athena RC}
 }
\email{nikos@cs.uoi.gr}

\author{Manolis Terrovitis}
\affiliation{%
  \institution{Athena RC, Athens, Greece}
 }
\email{mter@athenarc.gr} 
\begin{abstract}
We propose \tree, an in-memory implementation of the
\bt that adopts the structure of the disk-based index (i.e., a
balanced, multiway tree), setting the node size to a memory block that
can be processed fast and in parallel using SIMD instructions.
A novel feature of the \tree is that it enables gaps (unused
positions) within nodes by duplicating key values.
This allows (i) branchless SIMD search within each
node, and (ii) branchless update operations in nodes without
key shifting.
We implement a frame of reference (FOR) compression mechanism, 
which allows nodes to have varying
capacities, and can greatly decrease the memory footprint of \tree.
We compare our approach to existing main-memory indices and learned
indices under different workloads of queries and updates and
demonstrate its robustness and superiority compared to previous work
in single- and multi-threaded processing.
\end{abstract}

\maketitle



\section{Introduction}
The \bt is the dominant indexing method for DBMSs, due to
  its low and guaranteed cost for (equality and range) query
  processing and updates.
  It was originally proposed as a disk-based index, where the objective is
  to minimize the I/O cost of operations.
  As memories become larger and cheaper,
  main-memory and hardware-specific implementations
  of the \bt \cite{BoyarL92,LehmanC86,RaoR99,RaoR00,BohannonMR01,KimCSSNKLBD10,
  fix2011accelerating,LevandoskiLS13a,Kaczmarski12,ShahvaraniJ16,YanLPZ19,KwonLNNPCM23},
  as well as alternative
  access methods for in-memory data \cite{Morrison68,AskitisS10,BohmSVFHL09,
    KissingerSHL12,MaoKM12,LeisK013,ZhangAPKMS16,BinnaZPSL18,ZhangLLAKKP18}
  have been proposed.
  The optimization objective in all these methods is minimizing the
  computational cost and cache misses during search.
  More recently,
  learned indices \cite{KraskaBCDP18,GalakatosMBFK19,FerraginaV20,
DingMYWDLZCGKLK20,ZhangG22,WuCYSKX22,KipfMRSKK020,WuZCCWX21,LiLZDYP23,ZhangQYB24}, which replace the inner nodes of the \bt by ML models have
been suggested as a way for reducing the memory footprint of indexing
and accelerating search at the same time.
  
In this paper, we propose \tree, a \bt for main memory data, which is
optimized for modern commodity hardware and data parallelism.
\tree adopts the structure of the disk-based
\bt (i.e., a balanced, multiway tree),
setting the node size to a 
memory block that can be processed in parallel.
At the heart of our proposal lies a data-parallel
{\em successor} operator ($succ$), implemented
using SIMD, which is applied at each tree level
for branching during search and updates and for locating the
search key position at the leaf level.
To facilitate fast updates, without affecting SIMD-based search,
we propose a novel implementation for gaps (unused positions) by
duplicating keys.
The main idea is that we write in each unused slot the next used
key value in the node or a global MAXKEY value if all subsequent slots
are unused. 
Our \tree  construction algorithm initializes sparse leaf nodes with
intentional gaps in them, in order to (i) delay possible splits and
(ii) reduce data shifting at insertions. 
Splitting also adds gaps proactively.
Finally, we apply a frame-of-reference (FOR) based compression method that 
allows nodes that use fixed-size memory blocks to have
{\em varying capacities}, which saves space and increases data parallelism.
The computational cost of search and update operations in \tree is
$O(\log_{f}n)$, where $f$ is the capacity of the nodes,
assuming that $f$ is selected such
that each node can be processed by a (small) constant
number of SIMD instructions.

\stitle{Novelty and contributions.} There already exist several
SIMD-based implementations of B-trees and k-ary search \cite{SchlegelGL09,
  KimCSSNKLBD10, ShahvaraniJ16, YanLPZ19, KwonLNNPCM23,Graefe24}.
In addition, updatable learned
indices \cite{DingMYWDLZCGKLK20} use gaps to facilitate fast updates.
Finally, key compression in B-trees has also been studied in previous
work \cite{BohannonMR01}. To our knowledge, our proposed \tree is the
first \bt implementation that gracefully combines all these features, achieving at the
same time minimal storage and high throughput.
In particular, the use of duplicate key vales in unused gaps/slots allows
(i) branchless, data-parallel SIMD search at each node, and (ii)
efficient key insertions and deletions with limited
shifting of keys within each node.
In addition, our compression scheme allows for direct application
of data-parallel search on compressed nodes.
To our knowledge, using duplicate keys in gaps within
each node for efficient data-parallel
search and updates at the same time is novel and has not been
supported by previous B-tree implementations \cite{Graefe24}.
We implement a version of optimistic lock coupling in \tree for
concurrency control and
extensively compare \tree with open-source
single- and multi-threaded
implementations of state-of-the-art non-learned and learned indices
on widely used real datasets, to find that \tree and its compressed
version consistently prevail in different workloads of reads and updates,
typically achieving 1.5x-2x higher throughput than the best competitor.

\stitle{Outline} Section \ref{sec:relatedwork} presents related
work. The \tree is described in Section \ref{sec:method} and its updates
and construction in Section \ref{sec:updates}. Section
\ref{sec:pkeys} describes \tree compression.
Implementation details and concurrency control are discussed in
Section \ref{sec:implementation} and \ref{sec:concurrency}, respectively. 
Section  \ref{sec:exp} includes our experimental evaluation. We conclude in Section \ref{sec:conclusions}.

\section{Related Work} \label{sec:relatedwork}

\subsection{B-tree}
The \bt is considered the de-facto access method for relational data, having substantial advantages over hash-based indexing with respect to construction cost, support of range queries, sorted data access, and concurrency control \cite{Graefe11,ElmasriN89, Graefe24}.
As memory sizes grow, the interest has shifted to in-memory access methods \cite{ZhangCOTZ15}.
Rao and Ross \cite{RaoR99} were the first to consider the impact of cache misses in memory-based data structures; they proposed
{\em Cache-Sensitive Search Trees} (CSS-trees), in which every node has the same size as the cache-line of the machine and does not need to keep pointers for the links between nodes, but offsets that can be calculated by arithmentic operations.
Rao and Ross \cite{RaoR00} also proposed the Cache Sensitive \bt (CSB+tree), which achieves cache performance close to CSS-Trees, while having the advantages of a \bt.
Chen et al. \cite{ChenGM01, ChenGMV02} showed how prefetching can significantly improve the performance of index structures by reducing memory access latency.
The pkB-tree  \cite{BohannonMR01} is an in-memory variant of the B-tree that uses partial-keys (fixed-size parts of keys), which
reduce cache misses and improve search performance. Zhou and Ross \cite{ZhouR03} investigated buffering techniques, based on fixed-size or variable-sized buffers, for memory index structures, aiming to avoid cache thrashing and to improve the performance of bulk lookup in relation to a sequence of single lookups.
Graefe and Larson \cite{GraefeL01} surveyed techniques that improve the perfomance of \bt by exploting CPU caches.

\subsection{(Data) parallelism in B-trees}
The advent of SIMD instructions and GPUs  
opened new perspectives for in-memory index structures.
In an early work, Zhou and Ross \cite{ZhouR02} explored the use of SIMD to parallelize key database operations (such as scans, joins, and filtering), minimizing branch mispredictions.
Schlegel et al. \cite{SchlegelGL09} present methods for SIMD-based k-ary search (find which out of k partitions contains a search key).
FAST \cite{KimCSSNKLBD10} optimizes k-way tree search by leveraging architecture-specific features such as cache locality, SIMD parallelism on CPUs, and massive parallelism on GPUs.
\cite{fix2011accelerating} introduced a ``braided'' \bt structure optimized for parallel searches on GPUs, enabling lock-free traversal using additional pointers. 
Kaczmarski \cite{Kaczmarski12} proposed
a bottom-up \bt construction and maintenance technique
using CPU and GPU for bulk-loading and updates.
Bw-Tree \cite{LevandoskiLS13a} is a highly scalable and latch-free \bt variant optimized for modern hardware platforms, including multi-core processors and flash storage.
Hybrid \bt \cite{ShahvaraniJ16} leverages both CPU and GPU resources to optimize in-memory indexing, by dynamically balancing the workload between the CPU and the GPU.
Yan et al. \cite{YanLPZ19} proposed a \bt tailored for GPU and SIMD architectures. This structure decouples the ``key region'', which contains keys of the \bt with the ``child region'', which is organized as a prefix-sum array and stores only each node’s first child index in the key region.
Kwon et al. \cite{KwonLNNPCM23} proposed DB+-tree, a \bt with partial keys, that utilizes SIMD and other sequential instructions for fast branching. PALM \cite{SewallCKSD11} is a parallel latch-free variant of the \bt, which is optimized for multi-core processors, enabling concurrent search and update operations.
Other works explore the implementation of B-trees on flash memory \cite{WuKC07,AgrawalGSDS09,LiH0LY10,JorgensenRSS11,NaLM12,AthanassoulisA14,JinYJYY16,WangZHZ20},  non-volatile memory \cite{ChenJ15,LiuCW20} and hardware transactional memory \cite{SiakavarasBNGK20}.

\subsection{Other in-memory access methods}
Besides B-trees, other data structures have also been used for in-memory indexing, especially trie-based ones, such as the HAT-trie \cite{AskitisS07,AskitisS10}, the generalized prefix tree (trie) \cite{BohmSVFHL09}, KISS-TREE \cite{KissingerSHL12}, and Masstree \cite{MaoKM12}.
Leis et al. \cite{LeisK013} proposed a fast and space-efficient in-memory
trie called ART, which dynamically adjusts its node sizes
providing a compact and cache-efficient representation.
ART uses lazy expansion and path compression to improve
space utilization and search performance.
Leis et al. proposed two synchronization protocols for ART in  \cite{LeisSK016}, which have good scalability despite relying on locks: optimistic lock coupling and the read-optimized write exclusion (ROWEX) protocol.
Height Optimized Trie (HOT) \cite{BinnaZPSL18} is an in-memory trie-based index  that reduces tree height through path compression and node merging.
SuRF (Succinct Range Filter) \cite{ZhangLLAKKP18}
leverages succinct tries to provide a space-efficient solution for range query filtering.

\subsection{Learned Indexing}
The advent of fast and accurate machine learning techniques inspired the design of a new type of index structure, called {\em learned index} \cite{KraskaBCDP18}.
The main idea is to learn a cumulative distribution function (CDF)
of the keys and define
a Recursive Model Index (RMI)
that replaces the inner nodes of a \bt by a hierarchy of models that 
can predict very fast the position of the search key.
FITing tree  \cite{GalakatosMBFK19} and PGM-index \cite{FerraginaV20} build upon RMI with a focus on improving model performance.
with provable worst-case bounds on query time and space usage.
The RadixSpline (RS) \cite{KipfMRSKK020} learned index can be constructed in a single traversal of sorted data. 
ALEX \cite{DingMYWDLZCGKLK20} is an updatable learned index structure, based on RMI.
ALEX utilizes a {\em gapped array} layout that gracefully distributes extra space between elements based on the model's predictions, enabling faster updates and lookups by exponential search.
Other updatable learned indices include CARMI \cite{ZhangG22},
NFL \cite{WuCYSKX22}, LIPP  \cite{WuZCCWX21}, and 
DILI \cite{LiLZDYP23}, and Hyper \cite{ZhangQYB24}.
Refs. \cite{MarcusKRSMK0K20,sosd-neurips,WongkhamLLZLW22} provide comprehensive evaluations on updatable learned indices and traditional indices including many important findings, based on tests on several real-world datasets.

\section{The \tree}\label{sec:method}


We propose \tree, an in-memory implementation of the \bt,
which supports fast searches and updates by exploiting data
parallelism (i.e., SIMD instructions).
We first present the data structure in Section \ref{sec:method:ds}.
Then, Section \ref{sec:succ} describes the implementation of the
successor operator applied to each node for branching and key location
during search and updates. 
Finally, Section \ref{sec:search} presents the algorithms for
equality and range search.
        
\subsection{The structure}\label{sec:method:ds}
\tree follows the structure of the \bt.
Each internal node of the tree fits up to $N$ references to nodes
at the lower level and up to $N-1$ keys.
Leaf nodes contain rid-key pairs, where a record-id (rid)
is the address (potentially on the disk) of the record that
has the corresponding key value.
We assume that keys are unique (if not, rid is replaced by a pointer
to a block that keeps the rid's of all records having
the corresponding search key).
The storage of the rid's is decoupled from the storage of the keys, i.e., they are stored in two different arrays, such that the rid array is accessed only if necessary (i.e., only if the key is found and we need access to the corresponding record).
Similarly, the storage of keys in an internal node is decoupled from
the storage of children addresses,
to facilitate fast search, as we explain later.
Each leaf node hosts the address of the next leaf in the total key order;
Leaf chaining efficiently supports range queries, where query results are in consecutive tree leaves.
For the \tree nodes we use a value of $N$ that facilitates fast and
parallel search, as we will explain in
Sections \ref{sec:pkeys} and \ref{sec:implementation}.  
For the efficient handling of updates, we allow gaps (i.e., unused
slots) in nodes, similarly to previous work
\cite{DingMYWDLZCGKLK20,WuZCCWX21,LiLZDYP23}, as will be discussed in
Section \ref{sec:updates}. 

\begin{figure}[h]
	\centering
\includegraphics[width=0.9\linewidth]{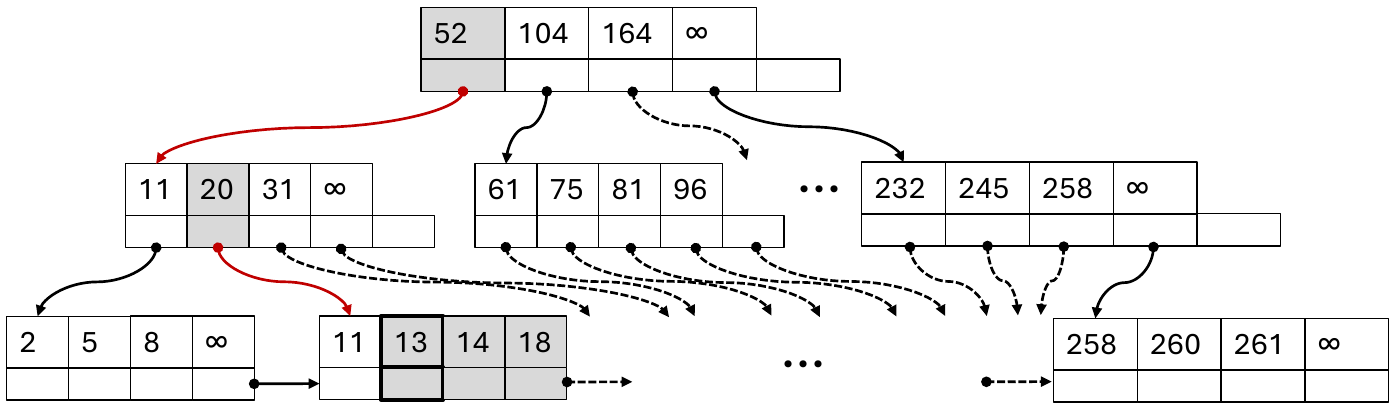}
	\caption{Example of \tree}
	\label{fig:bstree}
\end{figure}

Figure \ref{fig:bstree} shows an example of a \tree, where each node
holds up to $N-1=4$ keys. Each non-leaf node is shown as an array of
$N$ node pointers (bottom) and $N-1$ keys (top) that work as
separators. All keys in the subtree pointed by the $i$-th pointer are
strictly smaller than the $i$-th key and greater than or equal to the
$(i-1)$-th key (if $i>0$).
Any unused key slots at the end of each node carry a special \maxkey
value (denoted by $\infty$ in the figure), which is larger than the maximum
possible value in the key domain.
For example, if keys are unsigned 64-bit integers, \maxkey$=2^{64}-1$
and key values range in  $[0,2^{64}-2]$.

\vspace{-4mm}

\begin{figure*}
	\begin{tabular}{ccc}
          \includegraphics[width=0.27\linewidth]{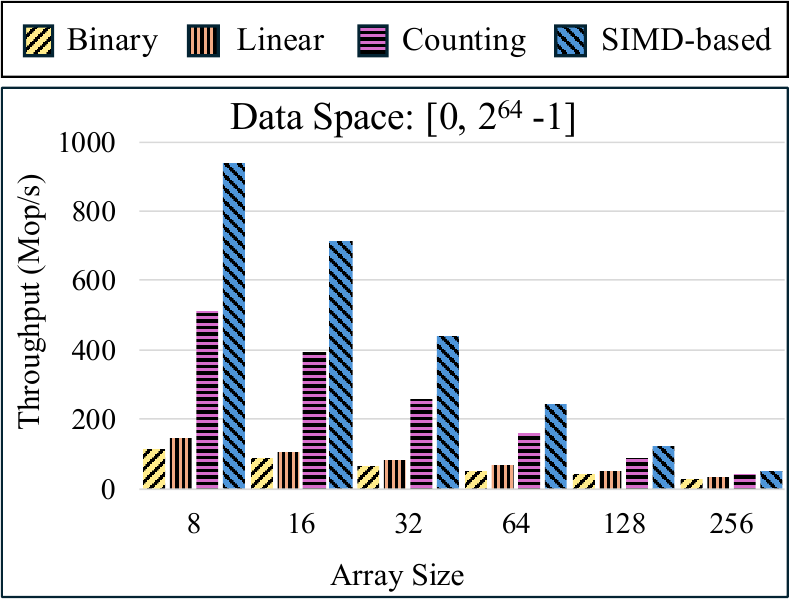}&
                                                                                                           \includegraphics[width=0.27\linewidth]{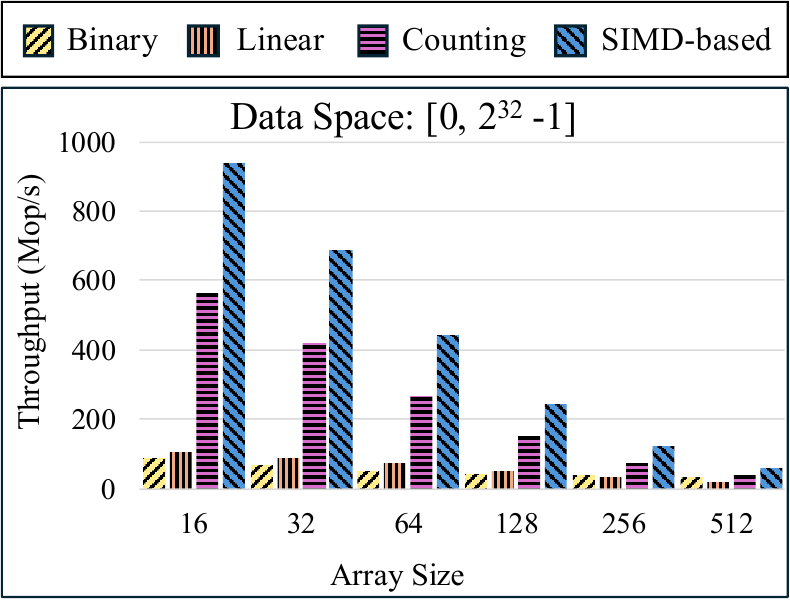}&
\includegraphics[width=0.27\linewidth]{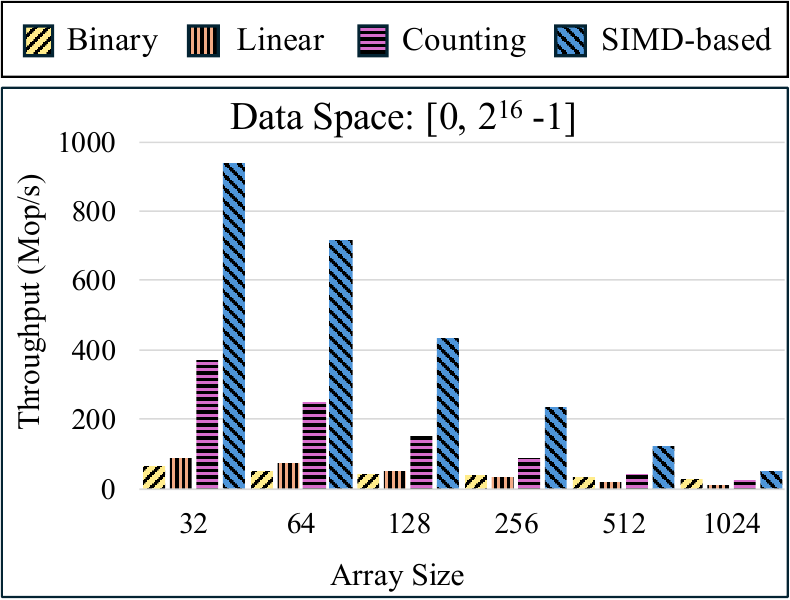}\\
          (a) \verb|uint64|  & (b) \verb|uint32|  & (c) \verb|uint16|\\
        \end{tabular}
        \vspace{-2mm}
      \caption{Comparison on successor search techniques in small \texttt{uint}
        arrays}
      \label{fig:searc_meth}
\end{figure*}

\subsection{Search within a \tree node}\label{sec:succ}
We now elaborate on the
implementation of {\em branching} at each node of the \tree, i.e., selecting the next node to visit.
Traditionally, at each visited node, starting from the root, finding the smallest key which is strictly greater that the query key $k$ is done either by binary search or by linearly scanning the entries until we find the first key greater than $k$.
Both these approaches incur significant CPU cost due to branch mispredictions.

We use an efficient implementation of the {\em successor}
operator applied to each node along the search path,
which does not involve search decisions. We denote by $succ_>$
the operator that finds the {\em smallest key position} which is {\em
  strictly greater than} the query key $k$ (used in non-leaf nodes) and by $succ_\ge$ the
finding of the {\em smallest key position} which is {\em greater than or equal to} $k$
(used in leaf nodes).
For example, in Figure \ref{fig:bstree}, $succ_{>124}(root)=2$,
as the smallest key which is greater than 124 is at position 2.
In general, $succ_>$ is the most frequently applied operation during
search, as we apply it in each non-leaf node along the path from the root to the leaf that includes the (first) search result.

Our approach  exploits data parallelism (i.e., SIMD instructions)
and does not include if statements or while statements
with an uncertain number of loops.
Specifically, let $v$ be a node and $k$ be the search key.
Then, $succ_{>k}(v) = |\{x: x \in v.keys \land  k\ge x\}|$, where
$|S|$ denotes cardinality of $S$.
Based on this, $succ_{>k}(v)$, for \verb|uint64| keys can be implemented by
code Snippet \ref{algo:counting_search}. The corresponding SIMD-fied
code (AVX 512) is Snippet \ref{algo:SIMD_search}, where Line 6 loads
the node keys
vector, Line 7 creates a comparison mask which has 1 at key positions
where the search key is greater than the node key,
and Line 8 counts the 1's in the mask.
As the code snippets do not include if-statements, they do not incur branch mispredictions.
CAPACITY is the (fixed) maximum capacity of the node,
so the number of iterations of the for-loop is hardwired;
all these favoring data parallelism.

%
%

\begin{algorithm}
	\renewcommand{\algorithmcfname}{Snippet}  
	\caption{Counting search}
	\label{algo:counting_search}
	\footnotesize
	\DontPrintSemicolon
	\texttt{int succG(Node *v, uint64 skey) \{ }\;
	\hspace{12pt}\texttt{int count = 0;\vspace{2pt}}\;
	\hspace{12pt}\texttt{for(int i=0; i<CAPACITY; i++) }\;
	\hspace{24pt}\texttt{count +=  (skey >= v->keys[i]);}\;
	\hspace{12pt}\texttt{return count;}\;
	\texttt{\} }\;
\end{algorithm}
\vspace{-4mm}

\begin{algorithm}
	\renewcommand{\algorithmcfname}{Snippet}  
	
	\caption{SIMD-based counting search (AVX 512)}
		\label{algo:SIMD_search}
	\footnotesize
	\DontPrintSemicolon
	\texttt{int succG\_SIMD(Node *v, uint64 skey) \{ }\;
	\hspace{12pt}\texttt{int count = 0;}\;
	\hspace{12pt}\texttt{\_\_mmask8 cmp\_mask = 0;}\;
	\hspace{12pt}\texttt{\_\_512i vec,Vskey = \_mm512\_set1\_epi64(skey);\vspace{2pt}}\;
	\hspace{12pt}\texttt{for(int i = 0; i < CAPACITY; i += 8) \{ }\;
	\hspace{24pt}\texttt{vec = \_mm512\_loadu\_epi64((\_\_512i*)(v->keys+i));}\;
	\hspace{24pt}\texttt{cmp\_mask = \_mm512\_cmpge\_epu64\_mask(Vskey,\hspace{1pt}vec);}\;
	\hspace{24pt}\texttt{count +=  \_mm\_popcnt\_u32((uint32\_t)\hspace{1pt}cmp\_mask);}\;
	\hspace{12pt}\texttt{\} }\;
	\hspace{12pt}\texttt{return count;}\;
	\texttt{\} }\;
\end{algorithm}

Similarly,
$succ_{\ge k}(v) = |\{x: x \in v.keys \land  k> x\}|$
and the same code snippets can be used, by replacing  comparison
\verb|>=| by \verb|>| and 
\verb|_mm512_cmpge_epu64_mask| by \verb|_mm512_cmpgt_epu64_mask|.
This approach (with slightly different implementation) has also been
suggested for SIMD-based k-way search in \cite{ShahvaraniJ16,ZhouR02,SchlegelGL09}.

The experiments of Figure \ref{fig:searc_meth}
illustrate the
efficiency of data-parallel $succ_{>}$ compared to traditional
implementations of
branching in multiway trees, on sorted arrays of 64-bit,  32-bit, and
16-bit unsigned integers.
The arrays simulate
key arrays in a full \tree node, with values drawn randomly from the
corresponding \verb|uint| domain.
We performed random successor (i.e., branching)
operations to the array and measured the throughput (in millions
operations per second) of four
implementations%
\footnote{See Section \ref{sec:exp} for our
experimental setup.}
of $succ_{>}$:
\begin{itemize}
\item {\bf Binary:} use of (non-recursive) binary search
\item {\bf Linear:} scan data from the beginning until 
  successor is found
\item {\bf Counting:} count-based in a for-loop (Snippet
  \ref{algo:counting_search})
\item {\bf SIMD-based:} count-based using SIMD (Snippet \ref{algo:SIMD_search})
\end{itemize}
We tested various sizes of the array, modeling different key-array
sizes in a \tree node.
We used array sizes that are multiples of 8, which allows us to take
full advantage of SIMD-parallelism.

Binary search and linear scan perform similarly, with linear search
being superior on small arrays
and binary search prevailing on larger arrays, as expected.
Counting search (Snippet \ref{algo:counting_search})
is much faster than binary/linear scan, 
due to the absence of branch instructions and due to
autovectorization optimizations at the assembly level,
by the \verb|-O3 -march=native| compilation flag.

Observe the excellent performance of SIMD-based
search (Snippet \ref{algo:SIMD_search})
for all array and key sizes.
Compared to
black-box \verb|-O3| compilation,
custom vectorization offers significant
advantages and achieves the theoretically optimal performance.
For example, for 64-bit keys and key-array capacity 16, it
achieves 7x performance improvement over binary search,
which is even higher than the theoretically expected 4x
($\log_{16}16$ vs. $\log_216$).
For \verb|uint16| keys,
Snippet \ref{algo:SIMD_search}
is more than 2x faster than
autovectorized Snippet \ref{algo:counting_search}.









	


\subsection{\tree search}\label{sec:search}

Algorithms \ref{algo:E_search} and \ref{algo:R_search} show how the \tree is searched for (i) equality queries and (ii) range queries.
For equality, we traverse the tree by applying a $succ_{> k}(v)$
operation at each non-leaf node $v$. At the reached leaf $v$ we apply
a $succ_{\ge k}(v)$ operation
to find the first position $r$ in the leaf having a key greater than or equal to $k$.
If $v.keys[r]$ equals $k$, then the record at position
$r$ is returned; otherwise, $k$ does not exist.
Equality search requires
one $succ_{>}$ or $succ_{\ge}$ operation per node along the search path. 

For range queries, assume that we are looking for all keys $x$, such
that $k_1\le x \le k_2$.
We traverse the tree using $succ_{> k_1}$ operations to find the first leaf that may contain a query result. 
In that leaf, we apply a $succ_{\ge k_1}$ operation to locate
the position $r_1$ of the first qualifying key.
To find the end position $r_2$ of the query results, starting from the
current leaf $v$, we search for the leaf which includes the first key
greater than $k_2$, by performing one $succ_{> k_2}$ operation per
leaf.
Hence, range searches require one $succ_{>}$ operation per node
along the search path in search for $k_1$ plus one $succ_{>}$
operation for each leaf that includes range query results.
For large query ranges whose results appear in numerous leaves,
we apply an alternative implementation of
range queries, where one equality search is used to locate $r_1$ and
another equality search is then applied to locate $r_2$.
This is
expected to be faster than Algorithm \ref{algo:R_search} if the height
of the tree is smaller compared to the number of leaves that include
the query results.

\begin{algorithm}[t]
	\LinesNumbered
	\small
	\Input{search key $k$, \tree root node $v$}
	\Output{record-id corresponding to key $k$}
	\BlankLine
	\While{$n$ is non leaf}{
		$v \leftarrow $ node pointed by entry at position $v[succ_{> k}(v)]$
         }
         $r\leftarrow succ_{\ge k}(v)$ \comm*[f]{leaf node}\;
         \If{$v.keys[r]==k$}{
          {\bf return} record-id in $v$ at positon $r$
         }
         \Else($\triangleright$\ $k$ does not exist){ {\bf return} NULL}
	\caption{Equality Search}
	\label{algo:E_search}
      \end{algorithm}

\begin{algorithm}[t]
	\LinesNumbered
	\small
	\Input{search keys $k_1,k_2$, \tree root node $v$} 
	\Output{record-ids of keys $x$, where $k_1\le x \le k_2$}
	\BlankLine
	\While{$n$ is non leaf}{
		$v \leftarrow $ node pointed by entry at position $v[succ_{> k_1}(v)]$
         }
         $r_1\leftarrow succ_{\ge k_1}(v)]$\;
         \While{$(r_2\leftarrow v.keys[succ_{> k_2}(v)])==$ N}{
           $v \leftarrow nextleaf(v)$ \comm*[f]{next to $v$ leaf}\\
           \If{$v$ == NULL} {{\bf break} \comm*[f]{last leaf reached}}
         }
         {
           {\bf return} record-ids of keys from position $r_1$ to position $r_2$ (excl.)
          }

	\caption{Range Search}
	\label{algo:R_search}
      \end{algorithm}

As an example, consider searching for key $13$ in the tree of Figure
\ref{fig:bstree}. A $succ_{> 13}$ operation on the root will give
position 0, as there are 0 keys smaller than or equal to $13$, so the
first pointer of the root will be followed. Then, the $succ_{> 13}$
operation on the visited node will return 1, which means that we then
visit the 2nd leaf, where $succ_{\ge 13}$ is applied that returns 1,
i.e., the position of 13 in the leaf. A range search for keys in
$[13,17]$ first locates $13$ and
then finds the upper bound $18$ in the same leaf after applying  
$succ_{> 17}$.



\section{Gaps and Updates}\label{sec:updates}

	

The main novelty of \tree is the implementation of gaps in nodes using
{\em duplicated keys}, which facilitates efficient updates
and allows the use of Snippet 2 for search at the same time.
Specifically, while SIMD-based k-way search on {\em packed} arrays has
also been suggested before \cite{ShahvaraniJ16,ZhouR02,SchlegelGL09},
to our knowledge it has never been applied
on \bt nodes with unused slots.  
As discussed in Sec. \ref{sec:method:ds},
unused key slots at the end of each node 
are filled with \maxkey values; hence, uniqueness
is not a requirement for unused key positions. In addition, \tree does
not require the used key slots to be continuous at the beginning of
the node. This means that `gaps' with unused key slots are allowed in
a node.

\begin{figure}[h]
	\centering
	\includegraphics[width=\columnwidth]{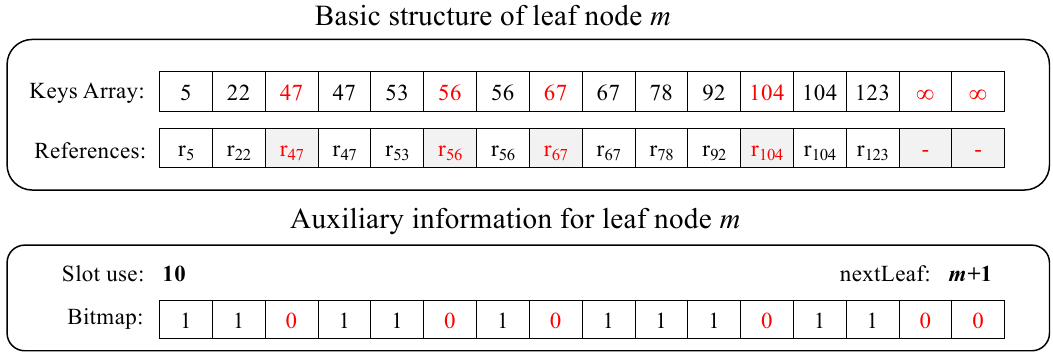}\\ 
\vspace{-2mm}
	\caption{\tree leaf node structure}
	\label{fig:example_structure}
\end{figure}

In \tree, we enforce that the key value
in a gap (unused slot) is the same as the {\em first subsequent non-gap key}.
By managing auxiliary information at each
node, i.e., (i) the {\em slot use} (number of used slots) and (ii) a
{\em bitmap} indicating used
slots,
we can efficiently
track unused slots;
see Figure \ref{fig:example_structure} for an example.
In the rest of this section, we will show how deletions and insertions are handled in the \tree. We will explain how our approaches minimize the overhead of node modifications while maintaining high search performance.



\vspace{-4mm}
\begin{figure}[h]
	\centering
	\includegraphics[width=\columnwidth]{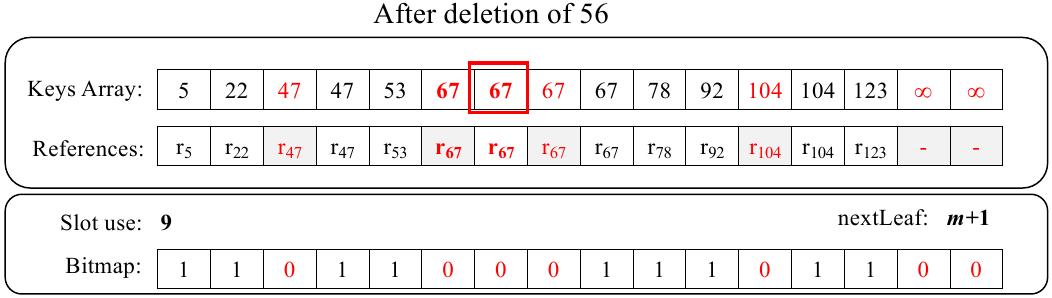}\\
        \vspace{0.3cm}
	\includegraphics[width=\columnwidth]{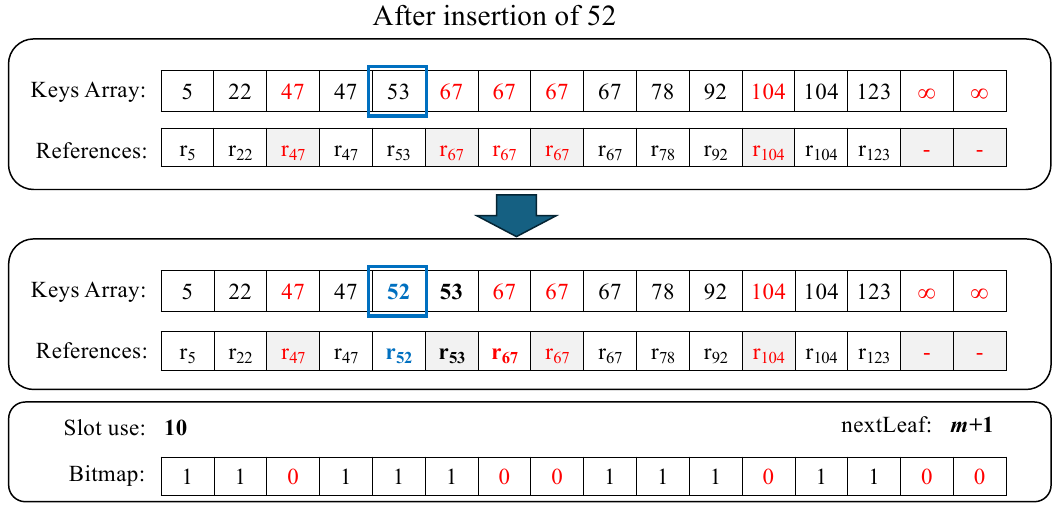}\\ 
\vspace{-2mm}
	\caption{Updates to \tree leaf node}
	\label{fig:updates}
\end{figure}
      
\subsection{Deletions} \label{sec:delete}
To delete a key, we first locate its position
$i$ in a leaf node, using the equality search
algorithm (discussed in Section \ref{sec:search}).
Then, we 
copy the key value from position $i+1$ to position $i$
and propagate it backwards
to previous gap positions in the node.
If $i$ is the last position in the leaf, we set
$v.keys[i]$=\maxkey.

One subtle point to note is that $succ_\ge$ may not give us the real
position of the key $k$ to be deleted but may give us the first of a
sequence of gaps that have key value equal to $k$.
For example, for deleting $k=56$ in Figure
\ref{fig:example_structure}, we apply $succ_\ge(56)$ which gives us
position 5. Then, we find the range of all positions having $56$
(i.e., [5,6]) and copy into them the next value (i.e., $67$).
Finding all the positions can be done very fast using bitwise operations.
Figure \ref{fig:updates} (top) shows the leaf node of
Figure \ref{fig:example_structure} after deleting key 56.
Algorithm \ref{algo:delete} is a pseudocode for deletions to the \tree.

As in previous work \cite{WuZCCWX21, RaoR00},
we do not take action for nodes with fewer
than 50\% occupied slots, as we anticipate insertions to be
more frequent than deletions,
so node merges or key
redistributions are not expected to pay-off.
If the last entry is deleted from a node, the node is marked as empty and
the corresponding separator entry at its parent is
`deleted' by copying the next key into it.

\begin{algorithm}[t]
	\LinesNumbered
	\small
	\Input{key $k$, \tree root node $v$}
	\Small
	\BlankLine
	 find leaf $v$ and position $r$ by running lines 1-3 of Alg \ref{algo:E_search}
	
	\If{$v.keys[r] \ne k$}{
		\bf return FAIL
	}
	\BlankLine
	$bitmap\leftarrow v.bitmap$\\
	
	\If(\comm*[f]{last key in node}){$r == N-1$}{
		$bitmap\leftarrow bitmap \oplus$ \texttt{0x0001}\\
		$replicasOfKey\leftarrow \_tzcnt(bitmap)$
		\BlankLine
		\For(\comm*[f]{propagate backwards}){$i \gets 0$ \textbf{to} $replicasOfKey-1$}{
			$v.keys[r-i]\leftarrow \maxkey$ 
		}
              }
              	\Else(\comm*[f]{$r$ is not the last position in the leaf}) {  	
		$bitmap\leftarrow bitmap \oplus ($\texttt{0x8000} $\ll r)$\\
		$replicasOfKey\leftarrow \_lzcnt(bitmap)$\\
		$nextValidKey\leftarrow v.keys[r+replicasOfKey+1]$
		\BlankLine
		\For(\comm*[f]{propagate backwards}){$i \gets 0$ \textbf{to} $replicasOfKey$}{
			$v.keys[r+i]\leftarrow nextValidKey$ 
		}
		\BlankLine
		$bitmap\leftarrow bitmap \oplus (0x8000 \gg (r + replicasOfKey))$
		
	}
	
	$v.slotuse\leftarrow v.slotuse-1$\\
	$v.bitmap\leftarrow bitmap$\\
	\bf return SUCCESS
\caption{Deletion in \tree}
\label{algo:delete}
\end{algorithm}


\subsection{Insertions}\label{sec:insert}
Inserting a new key $k$ to the \tree entails searching for the leaf node
and the position in it to place it. Search is conducted by applying
$succ_{>k}$ operations starting from the root and following the
corresponding pointers. When we reach the leaf $v$ where in $k$ should be
inserted, we apply a $succ_{\ge k}$ operation which finds the
proper slot in $v$ to insert $k$.
Then, we verify whether the slot $i$ returned by
$succ_{\ge k}$ is occupied by another key. This is done by a simple
test. If $v.keys[i]=v.keys[i+1]$, then we are sure that position
$i$ is free, so we place $k$ there and finish.
For example, assume that we want to insert key 55
to the leaf node of
Figure \ref{fig:example_structure}.
We apply a $succ_{\ge k}$ operation to the leaf, which will give us
position 5.
Since the next position (6) has the same key, position 5 corresponds
to a free slot (gap), hence, we have directly put the inserted key 55 there.  
On the other hand, if the key at position $i$ is different compared to
the key at position $i+1$, this means that the position is
occupied.
In this case, we first find the first position $j$ after $i$,
which is unused (i.e., a gap), and right-shift all keys (and the
corresponding record pointers) from position $i$ to position $j-1$, to
make space, so that key $k$ can be inserted at position $i$. If there
is no free position after $i$, then we move one position to the left
(left-shift) all keys and record ids from position $i$ until the first
free position to the left of $i$.
Figure \ref{fig:updates} (bottom) shows an example of inserting key
52.
As the slot where 52 should go is occupied by 53 and it is not a gap
(the key following 53 is not equal to 53), we search for the next gap,
which is the position next to 53, right-shift 53 there and make room
for the new key 52. Algorithm \ref{algo:insert} describes \tree insertion.

In case leaf $v$ is full, then we conduct a {\em split} of $v$ and
introduce a new leaf node. The existing keys in $v$ together with $k$
are split in half and distributed between the two leaves. Instead of
placing the distributed keys to the first half of each of the two leaves, we
interleave each key with a gap to facilitate fast insertion of future keys.
Proactive gapping is
described in the next subsection.

\begin{algorithm}[t]
	\LinesNumbered
	\small
	\Input{key $k$, \tree root node $v$}
	\Small
	\BlankLine
	compute leaf $v$ and position $r$ by running lines 1-3 of Alg \ref{algo:E_search}
	
	\If{$v.slotuse < N$}{
		$bitmap\leftarrow \hspace{4pt}\neg v.bitmap$\\
		\If(\comm*[f]{$r$ is an empty slot}){$v.keys[r] == v.keys[r+1]$}{
			$v.keys[r]\leftarrow k$\\
			$bitmap\leftarrow bitmap \oplus (0x8000 \gg r)$\\
		}
		\Else{
			$keysForShift\leftarrow \_lzcnt(bitmap \ll r)$\\ 
			
			\If(\comm*[f]{empty slot to the right}){$keysForShift < N$}{
				shift right $keysForShift$ keys of node $v$\\
				$v.keys[r]\leftarrow k$\\
				$bitmap\leftarrow bitmap \oplus (0x8000 \gg(r + keysForShift))$\\
			}
			\Else(\comm*[f]{empty slot to the left}){
				\scriptsize$keysForShift\leftarrow \_tzcnt(bitmap \gg (N-r-1)) - 1$\\
				\Small shift left $keysForShift$ keys of node $v$\\
				$v.keys[r-1]\leftarrow k$\\
				\scriptsize$bitmap\leftarrow bitmap \oplus (0x8000 \gg(r - (keysForShift + 1)))$\\
			}
			
			\Small$v.slotuse\leftarrow slotuse + 1$\\
			$v.bitmap\leftarrow \hspace{4pt}\neg bitmap$\\
		}
	}
	\Else{
		split leaf node $v$
	}
		
	\bf return
	\caption{Insertion in \tree}
	\label{algo:insert}
\end{algorithm}

\subsection{Tree building}\label{sec:construction}
We now describe the algorithm for building a \tree from a set of keys
(bulk loading). Like typical \bt construction algorithms, we first
sort the keys to construct the leaf level of the index. To facilitate
fast future insertions, we do not pack the nodes with keys, that is
we leave free space to accomodate future insertions.
Specifically, if $N$ is the capacity of a leaf node, we construct all
leaf nodes by adding to them the keys in sorted order;
each leaf node takes 
$\alpha\cdot N$ (key, record-id) pairs, where $\alpha$ ranges from $0.5$
(half-full nodes) to $1$ (full nodes).
We typically set $\alpha=0.75$ for tree leaves.
For each leaf, instead of placing all keys at the beginning of the
leaf and leaving $(1-\alpha) N$ consecutive empty slots at the end of
the node, we {\em spread} the entries in the leaf by placing one gap
(empty slot) after every $\frac{1}{1-a} - 1$ entries.%
\footnote{Gaps between consecutive key values (for integer keys) are not introduced.}
For each leaf node (except the first one) a {\em separator key},
equal to the first key of the leaf, is added to an array. For each
separator key, a node pointer to the previous leaf is associated to
the separator. Finally, a node pointer to the last leaf is introduced
at the end of the array (without a key value). After constructing the
leaves, the (already sorted) array of separator keys is used to
construct the next level of \tree (above the leaves), recursively.
We increase $\alpha$ as we go up, since we anticipate
much fewer insertions (and node overflows) at higher levels.


\section{Key Compression}\label{sec:pkeys}
\tree , as it has been discussed so far, stores the exact keys in its nodes.
Previous work on key compression for \bt \cite{BohannonMR01} uses fixed-size partial keys. One issue with partial keys is the overhead of 
decompression which may compromise performance.
For \tree, we opt for the frame-of-reference (FOR) compression approach, which has minimal decompression overhead.

Specifically, for each node $v$, we store in the node's auxiliary information
(see Figure \ref{fig:example_structure})
the first key $v.k_{0}$ of the node and replace the $v$'s key array (of size $N$) by an array where each original key $k$ is replaced by the difference $k-k_{0}$.
This allows us to potentially double or quadruple the size of the array if the differences occupy much less space than the original keys. If $N=16$ and the original array stores 64 bits, it may potentially be replaced by an array of $N=32$ 32-bit differences or $N=64$ 16-bit differences.
Since the keys in a node are ordered, we expect the differences to be small, especially in leaf nodes, so the space savings due to the reduction in the number of nodes are expected to be significant.
To achieve optimal performance of our data-parallel $succ_>$ implementation, we set the key array size to 1024 bits, so $N$ can be 16, 32, or 64.
As we have seen in Fig. \ref{fig:searc_meth}, the cost of our SIMD-based
$succ_>$ (Snippet 2) on $N=64$ \verb|uint16| keys is the same as that on $N=16$ \verb|uint64| keys.
This means that, after compression, the height of the tree can decrease and search can be accelerated.

\stitle{Tree construction}
Our goal is to construct the tree in one pass over the sorted keys
and to result in leaf nodes having 75\% occupancy (except when we are dealing with regions of sequential key values), while achieving the best possible compression.
For this, we begin by checking whether the leaf can be filled with 16-bit differences for the keys. If this is not feasible, we reattempt the process by checking if half of the keys can be stored as 32-bit differences. If this attempt also fails, we conclude by storing the exact 64-bit keys.




\stitle{Search}
To apply $succ_{>k}$ at a node $v$ we first compute $k'= k-v.k_0$, where $v.k_0$ is the first key value of $v$, stored explicitly in $v$'s meta-data (this is the only decompression overhead).
Then, we apply $succ_{>k'}$ to the node to find the position of the node pointer to follow. 
The same procedure is applied at the leaf nodes for $succ_{\ge k}$; the position of $k'= k-v.k_0$ corresponds to the position of $k$, or if $succ_{\ge k'}$ returns NULL, $k$ does not exist.



\stitle{Insert} We can directly use the \tree insertion algorithm to insert a new key $k$, by first running the search algorithm discussed above to find the leaf $v$ and the position in $v$ whereto insert $k$ and then
store the difference $k-v.k_0$ there.
The new nodes after a splitting a node $v$ can be of the same type as $v$, or they can be further compressed as they include fewer entries than $v$ with the first and the last one having smaller differences. 

\stitle{Delete} Deletion is not affected by key compression. When the key to be deleted is found (represented exactly or by its difference to $k_0$) at the corresponding leaf node, we simply copy into it the value of the next key, or \maxkey if the deleted key is the last one in the leaf node.
In compressed nodes, \maxkey is the maximum value that can be represented using all available bits.
If the first key of a node is deleted, we do not change $k_0$ (as it is not stored in a slot of the array) and keep in the slots the differences to $k_0$.

\section{Implementation Details}\label{sec:implementation}
This section presents some implementation details of the \tree that have a significant impact in its performance in practice. 

\stitle{Node size and structure.} 
As in previous work \cite{SchlegelGL09,LevandoskiLS13a,ShahvaraniJ16,YanLPZ19,
ZhangAPKMS16,KwonLNNPCM23,LeisK013,BinnaZPSL18,ZhangLLAKKP18,DingMYWDLZCGKLK20,
WuZCCWX21,ZhangQYB24}, we aim at indexing large keys, each
being a 64-bit unsigned integer.
To take full advantage of our SIMD $succ_{>}$ implementation,
each node stores a maximum of $N=16$ entries;
based on this, we allocate $16 \times 64 = 1024$ bits for the keys of each node.
Hence, the keys of each node (internal or leaf) fill two cache lines (each cache line can store 64 bytes).
This means that for $succ_>$,
we perform 2 SIMD instructions per node by loading the keys at 2 registers of 512 bits (8 keys at each register).
$1024$ bits are also allocated for keys in the compressed \ctree nodes; a compressed key array may have 32 32-bit or 64 16-bit entries.

\stitle{Memory management.}
To store the \tree in memory, we utilize two main structures: one to store the inner nodes and another for the leaf nodes.
Each inner node consists of two arrays with 16 entries. The first array holds 64-bit keys, while the second contains 32-bit references to nodes.
32 bits are sufficient for the references because they are in fact offsets to fixed-length slots in memory arrays allocated for nodes
(one for inner nodes and one for leaf nodes).
The auxiliary data for each node are put in a separate dedicated array aligned with the node arrays.
Hence, each inner node has a size of 192 bytes, which fits into 3 cache lines.
Our tested \tree implementation only has keys and no record-ids
in its leaves, so a leaf node contains a single array of 16 64-bit keys, with each leaf node occupying 128 bytes, fitting into 2 cache lines.
The inner nodes are stored in a contiguous array, aligned to Transparent Huge Pages (2 MB) for efficiency.
The leaf nodes are also stored in a contiguous array, but their alignment depends on the array size.
If the array is smaller than 3 GB, we align it using huge pages. Otherwise, it is aligned per cache line (64 bytes).
Alignment plays a crucial role in optimizing both cache efficiency and the use of SIMD operations, making it a key factor in the performance of \tree.
By aligning data to cache lines, we minimize cache misses and ensure that the CPU can retrieve entire nodes in a single memory access, significantly speeding up operations.
By aligning inner nodes to huge pages (2 MB), we reduce translation lookaside buffer (TLB) misses.
We also make use of \_\_builtin\_prefetch, a compiler intrinsic that allows us to pre-load data into the cache before it is needed, reducing latency.
By combining cache-line and SIMD-friendly alignment, along with appropriate use of \_\_builtin\_prefetch, \tree
allows for SIMD acceleration, and reduces memory access latency, leading to significantly better overall performance.

\stitle{Compress or not?}
The compressed version of \tree with variable-capacity nodes (Section \ref{sec:pkeys}) may reduce the memory footprint of the index and improve its performance, but also comes with the overhead of explicitly keeping the first key of a node, which does not pay off for nodes having 64-bit differences that cannot be compressed.%
\footnote{As we want all leaf nodes to have the same fixed size (for alignment purposes), we do not allow the same \tree to have both uncompressed and compressed leaves.} 
Hence, we employ a {\em decision mechanism} for choosing between the construction of a \tree or a compressed \tree, based on the input data.
Before bulk-loading the tree, we virtually split the sorted keys input into segments of 13 keys each, subtract the smallest key from the largest key in each bucket, and calculate the number of leading zeros. After performing these calculations for all segments, we take the average number of leading zeros. If this average is greater or equal to 32 bits, we conclude that the dataset can benefit from a compact \tree compression, and we go ahead with its construction. Otherwise, we create a standard (uncompressed) \tree. The selection of 13 keys is not arbitrary, as we put 25\% gaps at each leaf, and the 13th key serves as the separator for the node.
We found out that compression is not effective for inner nodes, so our final compressed \tree implementation has uncompressed inner nodes and compressed leaves.  

\section{Concurrency control}\label{sec:concurrency}
A wide range of concurrency control techniques has been proposed for the \bt and other indices, including lock coupling \cite{BayerS77,Graefe11}, right-sibling pointers \cite{LehmanY81},
fine-grained locking with lock coupling and logical removals \cite{BronsonCCO10},  Bw-tree's lock-free  mechanism \cite{LevandoskiLS13a,WangPLLZKA18} and
Read-Optimized Write Exclusion (ROWEX) \cite{LeisSK016}.
For B-trees, Leis et al. \cite{LeisH019,LeisSK016}
proposed an Optimistic Lock Coupling (OLC) technique, which is easy to implement and highly efficient. In OLC, when a thread wants to read or modify a node, it first acquires an optimistic read lock, allowing it to traverse the tree while maintaining a local copy of the node's state. If the thread intends to perform an update, it checks whether the node has been modified by another thread since it was read. If not, it applies the changes and commits them atomically. In the event of a conflict (i.e., if another thread has modified the node), the thread rolls back its changes and retries the operation from the root of the tree.

Our current implementation of \tree employs the OLC mechanism \cite{LeisH019,LeisSK016} with a slight modification in how node splits are handled. In the original OLC, when a thread takes on a write task, it splits the first full node (inner or leaf) it encounters during traversal. After completing the split, the thread restarts its traversal until it finds a path where every node has at least one empty slot. This efficiently reorganizes node contents, ensuring that insertions triggering splits at multiple tree levels are managed without requiring a global lock.
The use of sparse nodes in \tree allows for a more flexible procedure. Specifically, when a split occurs at a leaf node, the restart process is only necessary if the separator must be stored in the first node of the couple, which has no available free slot. This relaxation reduces unnecessary re-traversals.

\section{Experiments}\label{sec:exp}
We experimentally compare \tree to alternative main-memory indices (learned and non-learned).
As in previous work \cite{WongkhamLLZLW22,BinnaZPSL18,LeisK013}, we have built and compared indices for key data only; record ids or references are not stored, but the objective of each index is to locate the position(s) of the searched key(s).
The implementation of all methods is in C++ and compiled with \verb|gcc| (v13) using the flags \verb|-O3| and \verb|-march=native|. The experiments were conducted on a system with an 11th Gen Intel® Core™ i7-11700K processor with 8 cores, running at 3.60 GHz, 128 GB of RAM, having AVX 512 support. The operating system used was Ubuntu 22.04. We extended the codebase of GRE \cite{WongkhamLLZLW22} to include \tree. 

\subsection{Setup}\label{sec:exp:setup}

\stitle{Datasets}.
We ran our tests on standard benchmarking real datasets used in previous work \cite{MarcusKRSMK0K20,sosd-neurips,WongkhamLLZLW22}; each on consists of unsigned 64-bit integer keys.
In Amazon BOOKS \cite{MarcusKRSMK0K20,sosd-neurips}, each key represents the popularity of a specific book.
In FB \cite{MarcusKRSMK0K20,sosd-neurips,SandtCP19}, each key is a Facebook user-id.
OSM \cite{MarcusKRSMK0K20,sosd-neurips}
contains unique integer-encoded locations from OpenStreetMap.
GENOME \cite{WongkhamLLZLW22,rao20143d} includes loci pairs from human chromosomes.
PLANET \cite{WongkhamLLZLW22,goo2017}, a planet-wide collection of integer-encoded geographic locations compiled by OpenStreetMap.
According to \cite{WongkhamLLZLW22,ZhangQYB24}, OSM, FB, GENOME, and PLANET are complex real-world datasets that can pose challenges for learned indices. In contrast, the key distribution of BOOKS is easy to learn. We did not conduct experiments using synthetic datasets with common distributions, as, according to \cite{MarcusKRSMK0K20}, it would be trivial for a learned index to model such distributions. 
Although we do not include experiments with non-integer keys,
\tree can be used for floats/doubles, by changing the SIMD intrinsics
in Snippet 2,
and for strings, after being dictionary encoded (see \cite{DingNAK20}).

\stitle{Competitors}.
We compare our proposed {\bf \tree} and its compressed version, denoted by {\bf \ctree}, with five updatable learned and non-learned indices, for which the code was publicly available by the authors.
We did not compare to methods found inferior in previous comparative studies \cite{BinnaZPSL18,WongkhamLLZLW22} and to those with proprietary or unavailable code (e.g., \cite{KwonLNNPCM23,ZhangQYB24}).

\stitle {\bf Non-learned Indices}.
STX library \cite{stx_code} is a fully optimized C++ implementation of a main-memory \bt.
We use the set-based implementation from STX, which does not store values in the leaf nodes. For its construction, we used its fast bulk-loading method. We used the default block size of STX (256 bytes), so each leaf node holds 32 keys ($32 \times 8 = 256$ bytes).
We used two versions of the STX tree: the first is the original code,
referred to as {\bf \bt}, while the second version creates 25\% empty
space at the end of each leaf node, denoted by {\bf Sparse~\bt} (for
fairness, as our  \tree also proactively introduces 25\%  of gaps in leaves).

We also compare to SIMD-based  implementations of HOT
\cite{hot_code,BinnaZPSL18} and ART \cite{LeisK013,art_code}.
These tries do not support bulk-loading. However, we found that pre-sorting the data improves the construction time for both and results in more efficient structures.
The HOT code release does not support range queries, so we
implemented them ourselves.
The ART code also lacks range query support,
but we were unable to implement it due to the trie's complex structure.
HOT cannot handle keys greater than  $2^{63}-1$, so we removed values exceeding this limit from certain datasets.


\stitle{Learned Indices}.
{\bf ALEX} \cite{alex_code,DingMYWDLZCGKLK20} and {\bf LIPP}
\cite{lipp_code,WuZCCWX21} are the state-of-the-art updatable learned
indices for key-value pairs \cite{WongkhamLLZLW22}.
To use them, for each dataset we used as value of each key the key itself.
ALEX and LIPP both support bulk-loading.
Neither uses SIMD intrinsics during search.
The reason is their large leaves (in the order
of KB), where SIMD-based search is not effective (see Fig.
\ref{fig:searc_meth}).
ALEX uses exponential search in leaves, while LIPP 
eliminates the last-mile search entirely by ensuring that predicted
positions are exact for each key in the index.
	
Note that all aforementioned \tree competitors do not utilize huge
pages, as this would not bring a significant impact on them.
In the implementations of these methods,
each node
occupies a random location in memory and connects to its child nodes
via pointers.
This design results in the retrieval of a significantly larger number
of memory pages into the TLB cache, thereby diminishing the advantages
that huge pages would otherwise offer.

\stitle{Workloads}.
We used several different workloads used to measure performance. First, we randomly selected 150 million entries from each dataset for the construction phase. where we sorted the data and applied bulk-loading (except for HOT and ART, which, however, both benefit from sorting).
For our workloads, we used 50 million keys, that are selected randomly (i.e., queries and updates hit a random region of the space).
Our workloads are:
\begin{itemize}
  \item {\bf Read-Only (Workload A)}: 100\% reads (equality searches).
  \item {\bf Write-Only (Workload B)}: 100\% writes (insertions).
  \item {\bf Read-Write (Workload C)}: 50\% reads, 50\% writes.
  \item {\bf  Range-Write (Workload D)}: 95\% range searches, 5\% writes.
  \item {\bf Mixed (Workload E)}: 60\% reads, 35\% writes, 5\% deletions.
\end{itemize}




\subsection{Construction Cost and Memory Footprint}

\begin{table}
	\caption{Construction time  (for 150 million keys)}
	\label{tab:construction_time}
	\centering
\vspace{-3mm}
	\resizebox{\columnwidth}{!}{%
		\begin{tabular}{|cccccc|}
			\hline
			\multicolumn{6}{|c|}{\textbf{Construction Time (sec)}}                                                                                                                    \\ \hline\hline
			\multicolumn{1}{|l|}{\textbf{Indices / Datasets}} &
			\multicolumn{1}{l|}{\textbf{BOOKS}} &
			\multicolumn{1}{l|}{\textbf{OSM}} &
			\multicolumn{1}{c|}{\textbf{FB}} &
			\multicolumn{1}{l|}{\textbf{GENOME}} &
			\multicolumn{1}{l|}{\textbf{PLANET}} \\ \hline
			\multicolumn{1}{|c|}{\tree} &
			\multicolumn{1}{c|}{\textbf{0.33}} &
			\multicolumn{1}{c|}{0.33} &
			\multicolumn{1}{c|}{0.33} &
			\multicolumn{1}{c|}{0.33} &
			0.33 \\ \hline
			\multicolumn{1}{|c|}{C\tree} &
			\multicolumn{1}{c|}{0.35} &
			\multicolumn{1}{c|}{\textbf{0.32}} &
			\multicolumn{1}{c|}{\textbf{0.18}} &
			\multicolumn{1}{c|}{\textbf{0.20}} &
			\textbf{0.18} \\ \hline
			\multicolumn{1}{|c|}{\bt}      & \multicolumn{1}{c|}{0.39}  & \multicolumn{1}{c|}{0.39}  & \multicolumn{1}{c|}{0.39}  & \multicolumn{1}{c|}{0.39}  & 0.39  \\ \hline
			\multicolumn{1}{|c|}{Sparse~\bt} & \multicolumn{1}{c|}{0.50}  & \multicolumn{1}{c|}{0.50}  & \multicolumn{1}{c|}{0.50}  & \multicolumn{1}{c|}{0.50}  & 0.50  \\ \hline
			\multicolumn{1}{|c|}{HOT}                     & \multicolumn{1}{c|}{15.61} & \multicolumn{1}{c|}{16.65} & \multicolumn{1}{c|}{16.56} & \multicolumn{1}{c|}{16.23} & 15.33 \\ \hline
			\multicolumn{1}{|c|}{ART}                    & \multicolumn{1}{c|}{5.65}   & \multicolumn{1}{c|}{6.11}   & \multicolumn{1}{c|}{6.62} & \multicolumn{1}{c|}{6.42} & 6.19 \\ \hline
			\multicolumn{1}{|c|}{ALEX}                    & \multicolumn{1}{c|}{25.43} & \multicolumn{1}{c|}{41.60} & \multicolumn{1}{c|}{45.46} & \multicolumn{1}{c|}{30.74} & 30.06 \\ \hline
			\multicolumn{1}{|c|}{LIPP}                    & \multicolumn{1}{c|}{9.58}  & \multicolumn{1}{c|}{9.31}  & \multicolumn{1}{c|}{6.98}  & \multicolumn{1}{c|}{7.01}  & 7.05  \\ \hline
		\end{tabular}%
	}
\end{table}

\begin{table}
	\caption{Memory footprint (for 150 million keys)}
	\label{tab:memory_footprint}
	\centering
\vspace{-3mm}
	\resizebox{\columnwidth}{!}{%
		\begin{tabular}{|cccccc|}
			\hline
			\multicolumn{6}{|c|}{\textbf{Memory Footprint (GB)}}                                                                                                                      \\ \hline\hline
			\multicolumn{1}{|l|}{\textbf{Indices / Datasets}} &
			\multicolumn{1}{l|}{\textbf{BOOKS}} &
			\multicolumn{1}{l|}{\textbf{OSM}} &
			\multicolumn{1}{c|}{\textbf{FB}} &
			\multicolumn{1}{l|}{\textbf{GENOME}} &
			\multicolumn{1}{l|}{\textbf{PLANET}} \\ \hline
			\multicolumn{1}{|c|}{\tree}    & \multicolumn{1}{c|}{1.84}  & \multicolumn{1}{c|}{1.84}  & \multicolumn{1}{c|}{1.84}  & \multicolumn{1}{c|}{1.84}  & 1.84  \\ \hline
			\multicolumn{1}{|c|}{C\tree} &
			\multicolumn{1}{c|}{2.03} &
			\multicolumn{1}{c|}{1.75} &
			\multicolumn{1}{c|}{\textbf{0.55}} &
			\multicolumn{1}{c|}{\textbf{0.80}} &
			\textbf{0.51} \\ \hline
			\multicolumn{1}{|c|}{\bt} &
			\multicolumn{1}{c|}{\textbf{1.41}} &
			\multicolumn{1}{c|}{\textbf{1.41}} &
			\multicolumn{1}{c|}{1.41} &
			\multicolumn{1}{c|}{1.41} &
			1.41 \\ \hline
			\multicolumn{1}{|c|}{Sparse \bt} & \multicolumn{1}{c|}{1.88}  & \multicolumn{1}{c|}{1.88}  & \multicolumn{1}{c|}{1.88}  & \multicolumn{1}{c|}{1.88}  & 1.88  \\ \hline
			\multicolumn{1}{|c|}{HOT}                     & \multicolumn{1}{c|}{1.78}  & \multicolumn{1}{c|}{1.79}  & \multicolumn{1}{c|}{1.83}  & \multicolumn{1}{c|}{1.92}  & 1.71  \\ \hline
			\multicolumn{1}{|c|}{ART}                    & \multicolumn{1}{c|}{7.23}   & \multicolumn{1}{c|}{7.36}   & \multicolumn{1}{c|}{7.42}  & \multicolumn{1}{c|}{7.38}  & 7.85  \\ \hline
			\multicolumn{1}{|c|}{ALEX}                    & \multicolumn{1}{c|}{2.73}  & \multicolumn{1}{c|}{2.77}  & \multicolumn{1}{c|}{2.77}  & \multicolumn{1}{c|}{2.73}  & 2.73  \\ \hline
			\multicolumn{1}{|c|}{LIPP}                    & \multicolumn{1}{c|}{13.51} & \multicolumn{1}{c|}{14.69} & \multicolumn{1}{c|}{10.89} & \multicolumn{1}{c|}{11.66} & 11.62 \\ \hline
		\end{tabular}%
	}
\end{table}
First, we compare all tested methods with respect to their
construction cost and memory footprint for 150M keys.
Since all indices require the data to be sorted (or benefit from sorting), we exclude sorting from the construction cost. Table 1 presents the construction times, while Table 2 shows the memory footprints. 
The construction time of our \tree also includes the decision-making mechanism (roughly takes 0.03 sec) on whether we will construct a \tree or a \ctree (see Section \ref{sec:implementation}).
To calculate the memory usage of each method, we utilize the C function \texttt{getrusage}%
\footnote{https://man7.org/linux/man-pages/man2/getrusage.2.html}.
Since all learned indices essentially store 64-bit values together
with the keys, we report half of their measured memory requirements,
to approximate the memory required just for the keys and their 
inner structure.

As expected,
non-learned indices (except from the \ctree) have a stable construction time and memory footprint.
On the other hand, the build cost of learned indices
is high and is affected by the data distribution;
their bottleneck is in training the models that
predict key positions.
The Sparse~\bt is larger than \bt, so it is costlier to build and
has a larger memory footprint.
Our \tree (and its compressed \ctree version) is faster to build compared to \bt  
and Sparse~\bt mainly due to our better memory management (static
pre-allocation vs. dynamic allocation) and because we use offset
addressing instead of memory pointers.

\ctree has the smallest memory footprint than all methods for FB, GENOME, and PLANET because of its high compression effectiveness, which also has a positive impact to the construction time.
On the other hand, \ctree occupies more space than \tree on BOOKS because the distribution of keys there does not provide many compression opportunities; in this case, most leaf nodes store 64-bit differences and also need to explicitly store the 64-bit key of the first key, which renders the size of the index even larger than that of \tree. Note that for BOOKS and OSM, our decision mechanism (see Section \ref{sec:implementation}) chooses to construct a \tree while for FB, GENOME, and PLANET it decides to construct a \ctree.

HOT is very expensive to build compared to \tree and
\bt, because HOT requires keys to be inserted one at a time, lacking a bulk-loading mechanism.
HOT uses slightly less memory than \tree; however, our \ctree has a significantly smaller memory footprint than HOT in three out of the five datasets.
ART is faster to build compared to HOT (but occupies more space),
due to its simpler insertion
process, however, it is also slow compared to B-trees for the same
reason as HOT.

Note that all  versions of \bt and \tree have smaller memory footprints compared to the learned indices.
LIPP is faster to construct compared to ALEX,
because it does not readjust the models when conflicts occur (two or more keys are mapped to the same position).
LIPP's downside is its large memory footprint,
caused by creating new nodes during conflicts,
whereas node reorganization in ALEX results in a smaller memory
footprint.

In conclusion, the \tree has low construction cost, with the \bt exhibiting comparable performance alongside a very small memory footprint. Additionally, the \ctree achieves the fastest construction time and consumes from 56\% to 94\% less memory than all methods in FB, GENOME, and PLANET. Our results align with the findings of Wongkham et al. \cite{WongkhamLLZLW22}, which conclude that memory efficiency is not a distinct advantage of updatable learned indices.

\subsection{Single-Threaded Throughput}






\begin{figure*}[h]
	\centering
	\includegraphics[width=\textwidth]{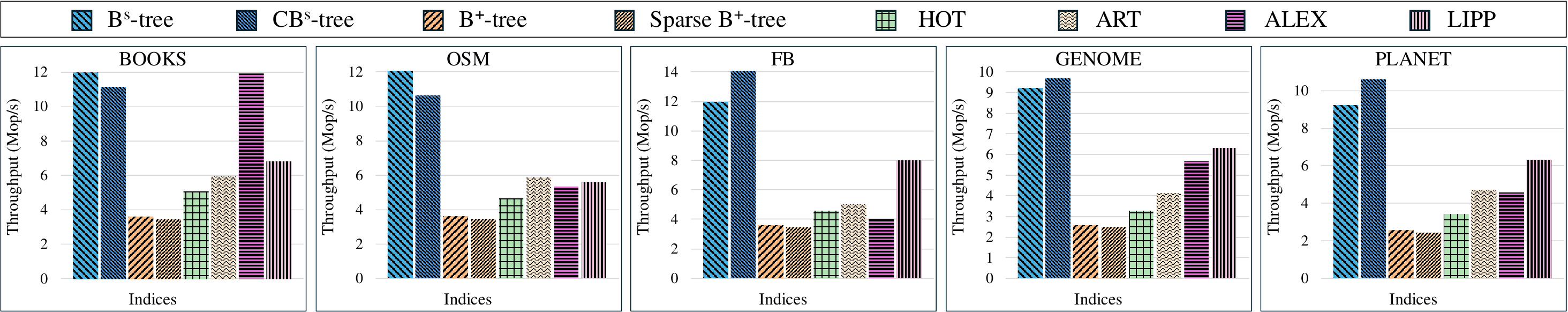}\\ 
	\vspace{-8pt}
	\caption{Workload A : Read Only (100\%)}
	\label{fig:s_thread_workload_A}
\end{figure*}
      
\begin{figure*}[h]
	\centering
	\includegraphics[width=\textwidth]{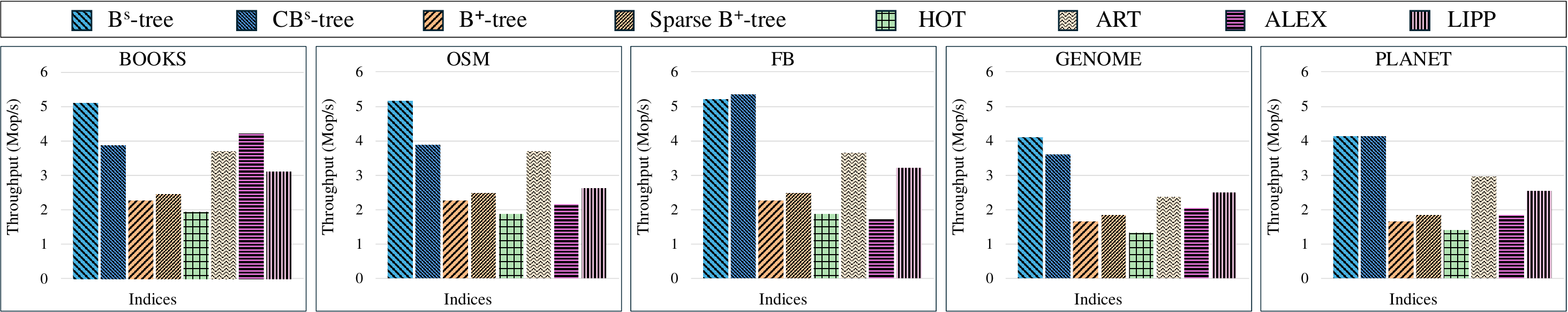}\\ 
	\vspace{-8pt}
	\caption{Workload B : Write Only (100\%)}
	\label{fig:s_thread_workload_B}
\end{figure*}

\begin{figure*}[h]
	\centering
	\includegraphics[width=\textwidth]{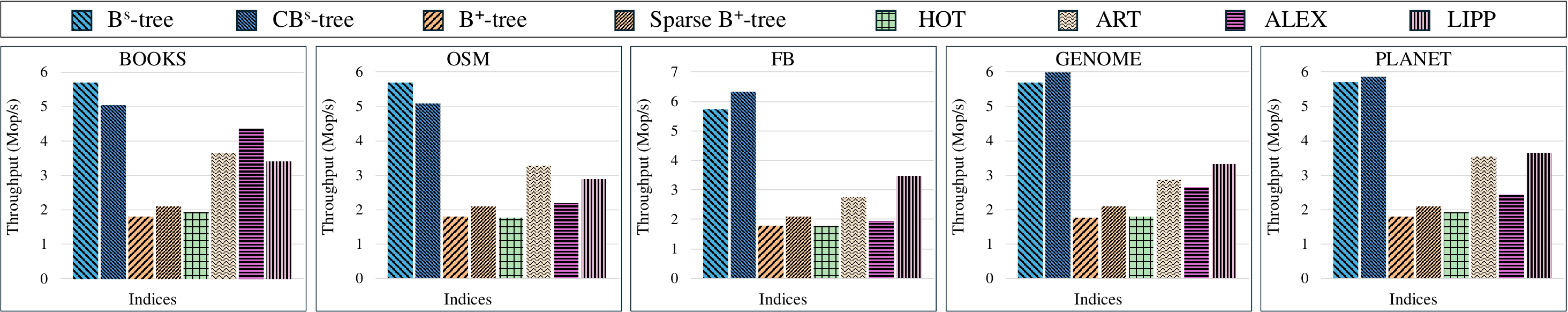}\\ 
	\vspace{-8pt}
	\caption{Workload C : Read (50\%) - Write (50\%)}
	\label{fig:s_thread_workload_C}
\end{figure*}

\begin{figure*}[h]
	\centering
	\includegraphics[width=\textwidth]{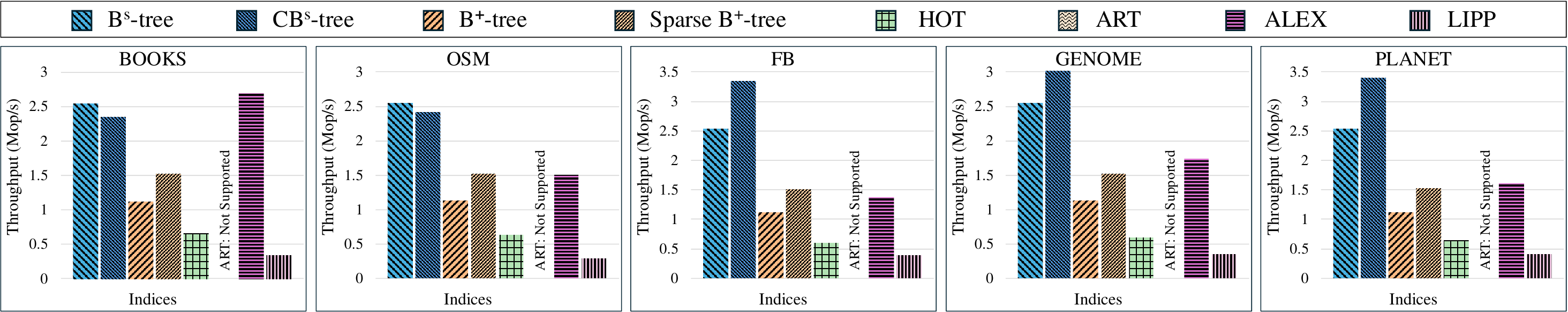}\\ 
	\vspace{-8pt}
	\caption{Workload D : Range (95\%) - Write (5\%)}
	\label{fig:s_thread_workload_D}
\end{figure*}

\begin{figure*}[h]
	\centering
	\includegraphics[width=\textwidth]{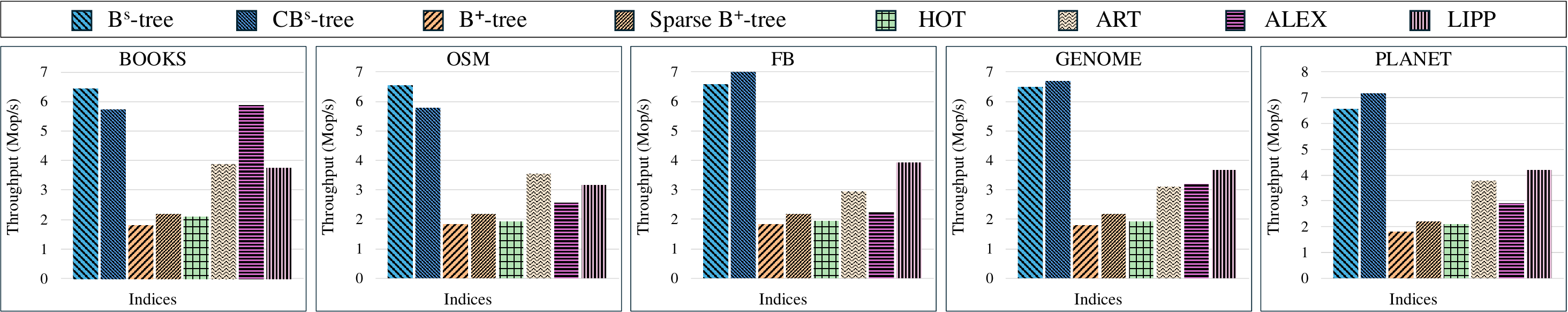}\\ 
	\vspace{-8pt}
	\caption{Workload E : Read (60\%) - Write (35\%) - Deletions (5\%)}
	\label{fig:s_thread_workload_E}
\end{figure*}

Next, we evaluate the throughput of all methods (single-threaded) on the five workloads described in Section \ref{sec:exp:setup}.
Figure \ref{fig:s_thread_workload_A} presents the throughput (millions of operations per second) of all methods for Workload A (read-only). \tree and \ctree outperform all competitors across the board except for BOOKS, where ALEX is marginally faster than \ctree and has the same throughput as \tree.
The excellent performance of ALEX on BOOKS is due to the smooth key
distribution there, which is easy for ALEX to learn.
On average, our methods have a significant performance gap compared to the nearest competitor.
Specifically, \tree is roughly 2.5x faster than HOT on OSM, 1.5x faster than LIPP on FB and GENOME and 2.2x faster than ART on FB and GENOME. 
\ctree is about 7\% slower than ALEX on BOOKS, but much faster than previous work on all other datasets and even faster than \tree on FB, GENOME, and PLANET, while having a much smaller memory footprint.
\ctree exploits the highly compressible keys of FB, GENOME, and PLANET
to drastically reduce the capacity of leaf nodes and the overall space
required for the index. This  increases the likelihood that multiple
searches hit the same leaves, exploiting the memory cache, as we will
also show in the next set of experiments.

For Workload B (write-only), as Figure \ref{fig:s_thread_workload_B}
shows, \tree outperforms all methods, while \ctree loses to ALEX only
on BOOKS. ART achieves relatively good performance in this workload
because of its lazy expansion and path compression.
LIPP is more robust than ALEX due to its accurate prediction mechanism, which reduces the need for frequent node splits or rebalancing due to the more appropriate placement of keys during its construction.
Observe that the Sparse \bt is faster than the \bt because it requires
much fewer splits. \ctree\ has competitive write performance compared
to \tree.

On Workload C (Figure \ref{fig:s_thread_workload_C}), 
the performance of all methods stands between that of Workload A (read-only) and Workload B (write-only), which is expected.
For the results of Workload D (range-write), see Figure~\ref{fig:s_thread_workload_D}.
The results are similar compared to Workload A (read-only) since Workload D is read-heavy.
Range queries retrieve 153 keys on average.
LIPP performs much worse than ALEX, as LIPP's structure is not optimized for range scans.
On the other hand, ALEX has large nodes and facilitates jumps to sibling nodes.
HOT is not optimized for range scans, so it performs poorly.
On the compressible datasets (FB, GENOME, PLANET), \ctree outperforms
\tree as its compressed leaves have larger capacities and fewer leaves
need to be scanned per query.
Finally, Figure~\ref{fig:s_thread_workload_E} shows the results using Workload E (read-write-delete).
The results are similar to those for Workloads A and B.
\tree and \ctree  outperform all competitors across all datasets except for BOOKS, where \ctree is slightly inferior to ALEX.
Deletions do not impose an overhead to \tree and all other methods.

\begin{figure*}[h]
	\centering
	\includegraphics[width=\textwidth]{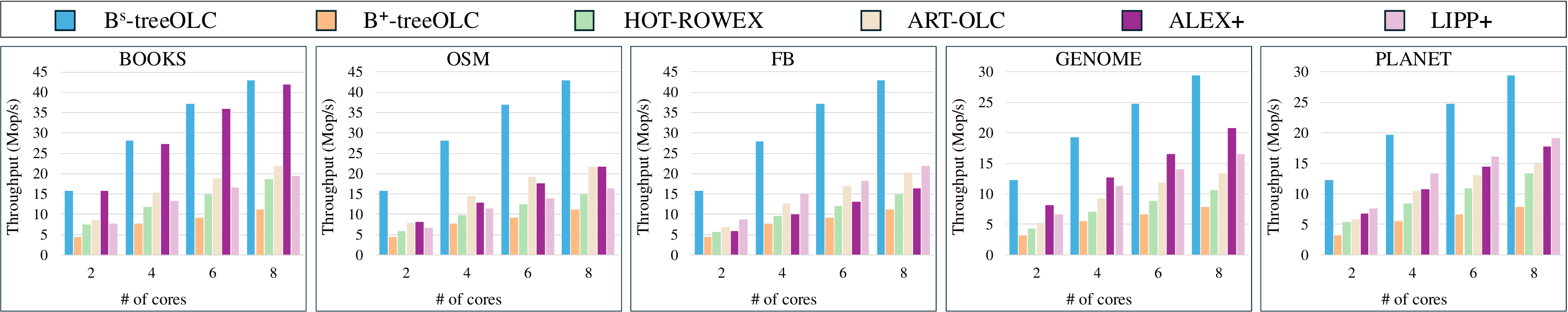}\\ 
	\vspace{-8pt}
	\caption{Workload A : Read Only (100\%) - OLC}
	\label{fig:m_thread_workload_A}
\end{figure*}

\begin{figure*}[h]
	\centering
	\includegraphics[width=\textwidth]{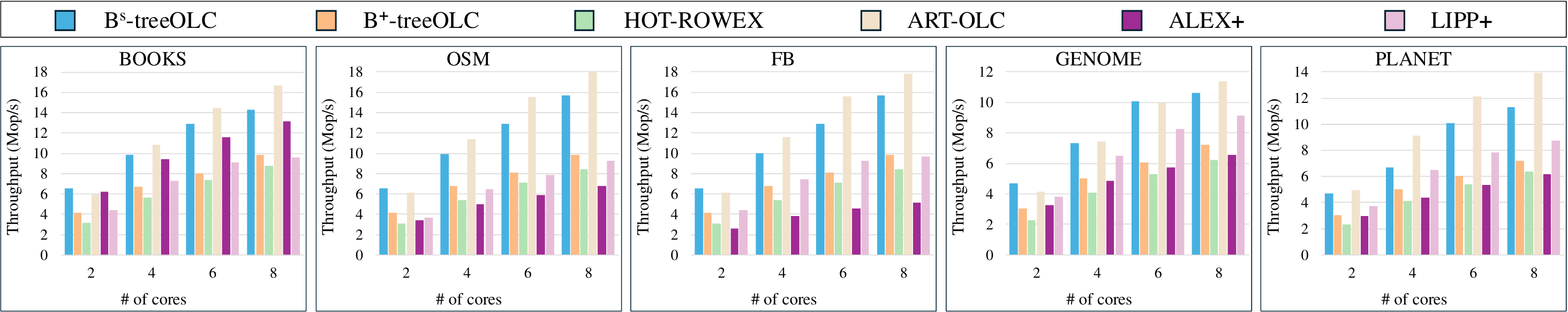}\\ 
	\vspace{-8pt}
	\caption{Workload B : Write Only (100\%) - OLC}
	\label{fig:m_thread_workload_B}
\end{figure*}

\begin{figure*}[h]
	\centering
	\includegraphics[width=\textwidth]{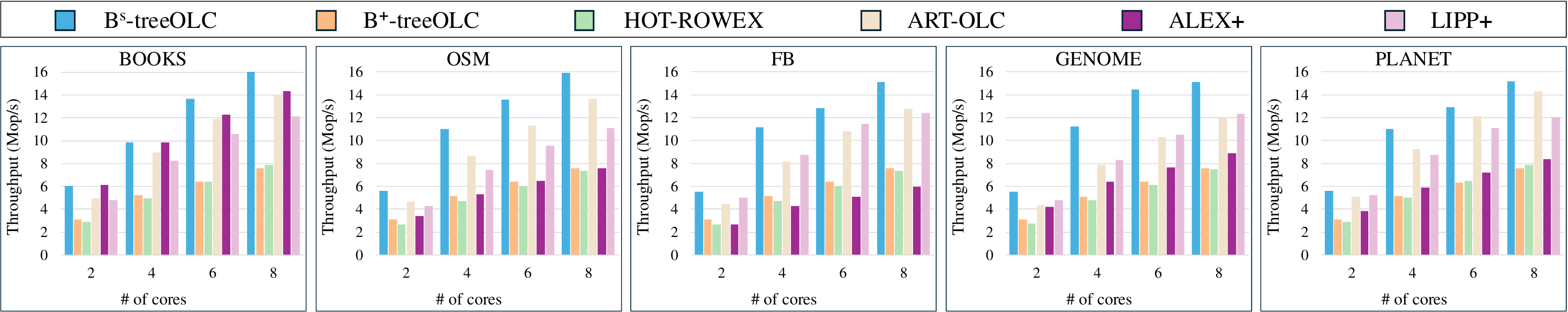}\\ 
	\vspace{-8pt}
	\caption{Workload C : Read (50\%) - Write (50\%) - OLC}
	\label{fig:m_thread_workload_C}
\end{figure*}

\subsection{Performance Counters}
Besides throughput, we also compared all methods with respect to various
performance counters, including instructions executed, cycles, 
mispredicted branches, and misses in L1 and TLB (Translation Lookaside Buffer) misses.
As representative datasets, we selected BOOKS and FB.%
\footnote{Recall that ALEX presents competitive behavior on BOOKS,
  whereas on FB \tree and \ctree have a large
  performance gap to other methods.}
We chose to measure Workload C (reads - writes),
as it is update-heavy and the 
unpredictable nature of writes can lead to numerous splits, stressing the indices.
To calculate these metrics, we utilized Leis's perf\_event code \cite{perfevent_code}.
The average performance measures per operation for all methods are presented in Table \ref{tab:perfCountersBOOKS_NEW}
and Table \ref{tab:perfCountersFB_NEW} for 
BOOKS and FB, respectively. 


\tree needs the smallest number of instructions and cycles on BOOKS,
with \ctree being a close runner up.
For FB, \ctree needs less cycles, which explains its superiority to \tree.
Our methods are simple and efficient, benefiting from the use of SIMD instructions and alignment,
which reduce the number of cycles and instructions required for each task.
They have the fewest mispredicted branches,
which is expected due to their branchless search;
mispredicted branches in \tree arise from the insertions.
Regarding L1, on average, our algorithms incur 16 cache misses, which is consistent with our expectation.
Specifically, the height of our trees is 6 and their fanout is 16; we encounter 2 cache misses per tree level and 2 cache misses at the leaf level ($2\times 6 + 2 = 14$ misses), with the remaining 2 misses caused by insertions.
For BOOKS, ART achieves the best performance in terms of L1 cache misses, though our algorithms are close in comparison.
Lastly, in terms of TLB misses, our algorithms exhibit outstanding performance relative to our competitors. This can be attributed to our use of huge pages (see Sec. \ref{sec:implementation}).
Overall, the performance counters show that our \tree and \ctree are
fully optimized and cache-efficient
via the use of SIMD instructions, huge pages, and
branchless code.

\begin{table}[]
	\centering
	\caption{Performance counters for BOOKS, Workload C}
	\label{tab:perfCountersBOOKS_NEW}
\vspace{-3mm}
	\resizebox{\columnwidth}{!}{%
		\begin{tabular}{|cccccc|}
			\hline
			\multicolumn{6}{|c|}{\textbf{Workload C - Dataset: BOOKS}}                                                                                                                          \\ \hline\hline
			\multicolumn{1}{|l|}{\textbf{Indices / Events}} &
			\multicolumn{1}{c|}{\textbf{Instr.}} &
			\multicolumn{1}{c|}{\textbf{Cycles}} &
			\multicolumn{1}{c|}{\textbf{Misp. Branches}} &
			\multicolumn{1}{c|}{\textbf{L1 Misses}} &
			\textbf{TLB Misses} \\ \hline
			\multicolumn{1}{|c|}{\tree}        & \multicolumn{1}{c|}{\textbf{220.18}}          & \multicolumn{1}{c|}{\textbf{884.16}}  & \multicolumn{1}{c|}{\textbf{1.03}}  & \multicolumn{1}{c|}{16.41} & 0.61 \\ \hline
			\multicolumn{1}{|c|}{C\tree}       & \multicolumn{1}{c|}{277.87}          & \multicolumn{1}{c|}{997.57}  & \multicolumn{1}{c|}{1.03}  & \multicolumn{1}{c|}{16.33} & \textbf{0.05} \\ \hline
			\multicolumn{1}{|c|}{\bt}        & \multicolumn{1}{c|}{656.84}          & \multicolumn{1}{c|}{2806.93} & \multicolumn{1}{c|}{11.49} & \multicolumn{1}{c|}{33.49} & 4.44 \\ \hline
			\multicolumn{1}{|c|}{Sparse \bt} & \multicolumn{1}{c|}{565.33}          & \multicolumn{1}{c|}{2408.02} & \multicolumn{1}{c|}{10.61} & \multicolumn{1}{c|}{28.29} & 3.84 \\ \hline
			\multicolumn{1}{|c|}{HOT}           & \multicolumn{1}{c|}{898.99} & \multicolumn{1}{c|}{2585.71} & \multicolumn{1}{c|}{3.44}  & \multicolumn{1}{c|}{31.78} & 4.75 \\ \hline
			\multicolumn{1}{|c|}{ART}           & \multicolumn{1}{c|}{435.56}          & \multicolumn{1}{c|}{1443.29} & \multicolumn{1}{c|}{2.12}  & \multicolumn{1}{c|}{\textbf{14.12}} & 3.06 \\ \hline
			\multicolumn{1}{|c|}{ALEX}          & \multicolumn{1}{c|}{612.57}          & \multicolumn{1}{c|}{1165.58} & \multicolumn{1}{c|}{5.23}  & \multicolumn{1}{c|}{21.03} & 2.22 \\ \hline
			\multicolumn{1}{|c|}{LIPP}          & \multicolumn{1}{c|}{300.47}          & \multicolumn{1}{c|}{1379.23} & \multicolumn{1}{c|}{1.95}  & \multicolumn{1}{c|}{18.66} & 5.02 \\ \hline
		\end{tabular}%
	}
\end{table}

\begin{table}[]
	\centering
	\caption{Performance counters for FB, Workload C}
	\label{tab:perfCountersFB_NEW}
\vspace{-3mm}
	\resizebox{\columnwidth}{!}{%
		\begin{tabular}{|cccccc|}
			\hline
			\multicolumn{6}{|c|}{\textbf{Workload C - Dataset: FB}}                                                                                                                             \\ \hline\hline
			\multicolumn{1}{|l|}{\textbf{Indices / Events}} &
			\multicolumn{1}{c|}{\textbf{Instr.}} &
			\multicolumn{1}{c|}{\textbf{Cycles}} &
			\multicolumn{1}{c|}{\textbf{Misp. Branches}} &
			\multicolumn{1}{c|}{\textbf{L1 Misses}} &
			\textbf{TLB Misses} \\ \hline
			\multicolumn{1}{|c|}{\tree}  & \multicolumn{1}{c|}{\textbf{220.51}}  & \multicolumn{1}{c|}{877.78}  & \multicolumn{1}{c|}{\textbf{1.03}}  & \multicolumn{1}{c|}{16.43} & 0.59 \\ \hline
			\multicolumn{1}{|c|}{C\tree} & \multicolumn{1}{c|}{269.33}  & \multicolumn{1}{c|}{\textbf{784.75}}  & \multicolumn{1}{c|}{1.09}  & \multicolumn{1}{c|}{\textbf{14.29}} & \textbf{0.00}  \\ \hline
			\multicolumn{1}{|c|}{\bt}    & \multicolumn{1}{c|}{655.42}  & \multicolumn{1}{c|}{2807.87} & \multicolumn{1}{c|}{11.54} & \multicolumn{1}{c|}{33.45} & 4.41 \\ \hline
			\multicolumn{1}{|c|}{Sparse \bt} &
			\multicolumn{1}{c|}{566.01} &
			\multicolumn{1}{c|}{2395.46} &
			\multicolumn{1}{c|}{10.62} &
			\multicolumn{1}{c|}{28.37} &
			3.85 \\ \hline
			\multicolumn{1}{|c|}{HOT}                   & \multicolumn{1}{c|}{962.58}  & \multicolumn{1}{c|}{2816.29} & \multicolumn{1}{c|}{4.08}  & \multicolumn{1}{c|}{33.47} & 5.06 \\ \hline
			\multicolumn{1}{|c|}{ART}                   & \multicolumn{1}{c|}{603.69}  & \multicolumn{1}{c|}{1804.94} & \multicolumn{1}{c|}{2.33}  & \multicolumn{1}{c|}{14.47} & 3.49 \\ \hline
			\multicolumn{1}{|c|}{ALEX}                  & \multicolumn{1}{c|}{1049.11} & \multicolumn{1}{c|}{2634.38} & \multicolumn{1}{c|}{6.97}  & \multicolumn{1}{c|}{40.31} & 4.72 \\ \hline
			\multicolumn{1}{|c|}{LIPP}                  & \multicolumn{1}{c|}{276.74}  & \multicolumn{1}{c|}{1350.58} & \multicolumn{1}{c|}{1.49}  & \multicolumn{1}{c|}{18.62} & 5.11 \\ \hline
		\end{tabular}%
	}
\end{table}

\begin{figure*}[h]
	\centering
	\includegraphics[width=\textwidth]{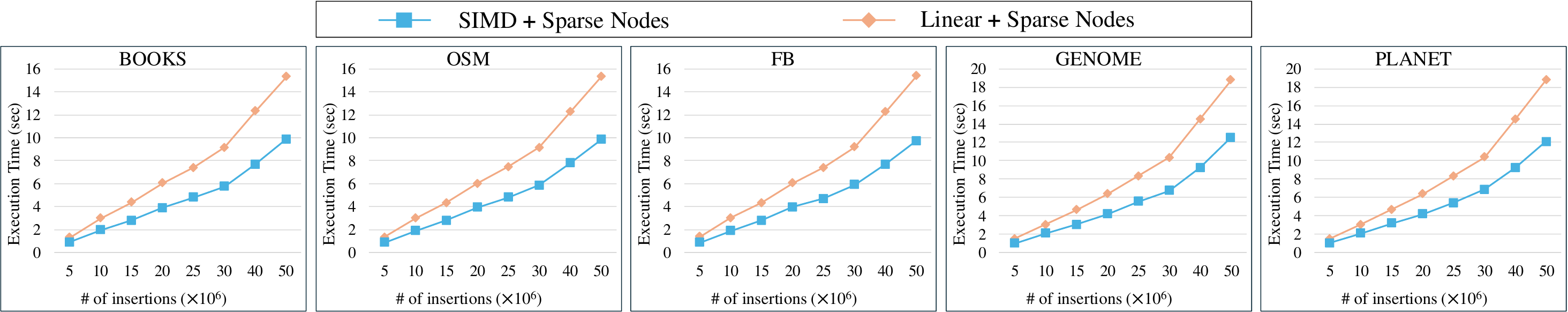}\\ 
	\vspace{-8pt}
	\caption{Impact of data parallelism to the Insertion cost of \tree}
	\label{fig:write_ratio}
\end{figure*}

\subsection{Multi-Threaded Workloads}

Next, we present multi-threaded experiments, where workloads are
executed by multiple threads simultaneously.
For concurrency control, our \tree uses OLC as described in Section
\ref{sec:concurrency}, ART and \bt use OLC, HOT uses ROWEX, and ALEX
and LIPP use optimistic locking. We applied Workload A, Workload B, and
Workload C, which are the most representative ones.
Figure~\ref{fig:m_thread_workload_A}, presents the throughput
(millions of operations per second) of all methods for Workload A
(read-only). Observe that the access methods exhibit the same behavior
as in single-threaded processing. Our \tree outperforms all competitors,
with ALEX being competitive only on BOOKS.
For the write-only Workload B (Figure~\ref{fig:m_thread_workload_B}),
ART achieves the best performance, while our \tree is competitive in
some cases. ART's strong performance is largely due to OLC, which is
well-suited to its structure.
For the mixed read-write Workload C
(Figure~\ref{fig:m_thread_workload_C}), our \tree outperforms all competitors across all datasets.
All methods scale well with the number of cores.

\subsection{Impact of \tree design}

In the final set of experiments, we assess the impact of the design choices in \tree. Figure \ref{fig:write_ratio} shows the impact of our data-parallel implementation using gaps that copy the key values from the next used slot,
on the cost of writes. We compare our \tree implementation to an
implementation of \tree, where nodes have gaps simply
supported by a bitmap (as in \cite{DingMYWDLZCGKLK20}).
In this implementation, branching in each node is done by linear scan
using the bitmap to ignore unused slots.%
\footnote{As nodes are small, linear scan is better than binary
  search. Besides, binary and exponential search require values in gaps
that obey the total order \cite{DingMYWDLZCGKLK20}.}
The figure shows that our SIMD implementation with gaps reduces the
cost of insertions (and node searches,
which are part of the insertion procedure) by 30\%-35\%.



\begin{figure}[h]
	\centering
	\includegraphics[width=0.7\linewidth]{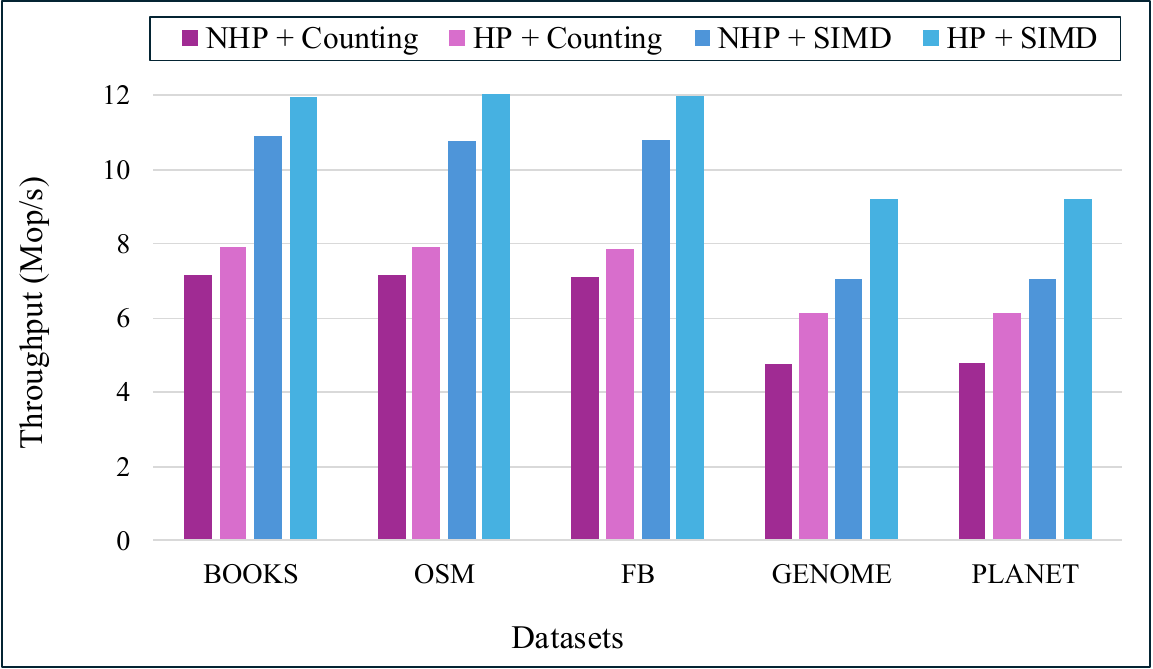}
	\caption{Effect of \tree design (Workload A)}
	\label{fig:opt_performance}
\end{figure}

We also analyze the effectiveness of the \tree features,
with the most powerful ones being SIMD-based branching (Snippet 2)
and transparent huge pages. We implemented four versions of our structure: one without any optimizations (NHP + Counting), one with only huge pages enabled (HP + Counting), another with only SIMD instructions enabled (NHP + SIMD), and finally, a fully optimized version that combines both huge pages and SIMD instructions (HP + SIMD).
Figure~\ref{fig:opt_performance} shows the throughput of the four versions of \tree for Workload A (reads). 
We observe that both optimizations (HP and SIMD) boost \tree.

\vspace{-2mm}
\subsection{Summary of Experimental Findings}
In summary, \tree and \ctree exhibit excellent and robust performance
for different workloads and different datasets of varying
distribution, being superior than all competitors in most cases, in
single- and multi-threaded processing.
They also have the lowest construction cost and memory footprint.
Note that our decision mechanism, which imposes a small overhead in
the construction (up to 10\% of the construction cost) decides
automatically and correctly which of \tree or \ctree to build for a
given dataset. \tree outperforms trie-based indices like HOT and ART,
except for multithreaded write-heavy workloads, where ART is
superior. Still, we have not implemented yet \tree for string keys, where
trie-based structures are known to perform best.
Regarding updatable learned indices, our study shows that they are
typically outperformed by optimized non-learned indices like our \tree.


\section{Conclusions}\label{sec:conclusions}
We proposed \tree, a  main-memory \bt
that uses a simple and intuitive data-parallel implementation for
branching at each level during search and updates. \tree is based on a novel implementation of gaps in nodes by
duplicating existing keys, which does not affect SIMD-based
branchless search in nodes. We use FOR compression mechanism
for nodes that include keys with small differences. Our
experimental evaluation demonstrates the superiority of \tree compared
to open-source state-of-the-art non-learned and learned indices, with
respect to construction time, memory footprint, and throughput for
various workloads that include queries and updates in single- and multi-threaded processing.
Inspired by \cite{ShahvaraniJ16}, in the future, we plan to implement \tree in a hybrid setup, where the top (and infrequently updated levels) are handled by the GPU that allows much higher data parallelism and the lower (frequently updated) levels are handled by the CPU.
In addition, we will study the support of string keys in \tree (one solution is to use Binary, ASCII or Base64 encoding \cite{DingMYWDLZCGKLK20}).

\section*{Availability}
The code of the \tree is publicly available at\linebreak
github.com/dTsitsigkos/BsTree-A-gapped-data-parallel-B-tree

\section*{Acknowledgements}
Work supported by project MIS 5154714
of the National Recovery and Resilience Plan Greece 2.0,
funded by the EU under the NextGenerationEU Program.

\bibliographystyle{ACM-Reference-Format}
\bibliography{references}

\end{document}